\documentclass[onecolumn,secnumarabic,amssymb, nobibnotes, aps, prd,showpacs,showkeys,unsortedaddress,10pt]{revtex4-1}

\usepackage{appendix}
\usepackage{epsfig}
\usepackage{graphicx}
\usepackage{color}
\usepackage{soul}
\usepackage{ulem}
\usepackage{amssymb}
\usepackage{hyperref}
\usepackage{multirow}
\usepackage{float}
\def\ee{\end{equation}}
\def\be{\begin{equation}}
\def\eea{\end{eqnarray}}
\def\bea{\begin{eqnarray}}
\def\bal{\begin{align}}
\def\eal{\end{align}}

\def\pbarp{{\bar p}p}

\def\energy8{$\sqrt{s}=8\ TeV$}
\def\energy14{$\sqrt{s}=14\ TeV$}

\def\.[{\big{[}}
\def\.[{\big{]}}
\def\.({\big{(}}
\def\.){\big{)}}

\begin{document}
\title{Elastic pp scattering  from the optical point to past the dip:
an empirical parametrization   from ISR to LHC }

\author{D. A. Fagundes}
\email{fagundes@ifi.unicamp.br}
\affiliation{INFN Frascati National Laboratories, Via E. Fermi 40, 00444, Italy}
\affiliation{Instituto de F\'{\i}sica Gleb Wataghin, Universidade Estadual de Campinas, UNICAMP, 13083-859 Campinas SP, Brazil}

\author{G. Pancheri}
\email{giulia.pancheri@lnf.infn.it}
\affiliation{INFN Frascati National Laboratories, Via E. Fermi 40, 00444, Italy}

\author{A. Grau}
\email{igrau@ugr.es}
\affiliation{Departamento de F\'{\i}sica Te\'orica y del Cosmos, Universidad de Granada, 18071 Granada, Spain}

\author{S. Pacetti}
\email{simone.pacetti@pg.infn.it}
\author{Y. N. Srivastava}
\email{yogendra.srivastava@pg.infn.it}
\affiliation{Physics Department and INFN, University of Perugia, 06123 Perugia, Italy}

\begin{abstract}
We describe the main features of recent LHC data on elastic $pp$ scattering
 through  a simple parametrization to the  amplitude, inspired by a model proposed by Barger and Phillips  in 1973, comprising of two exponentials with a relative phase. Despite its  simplicity, this parameterization  reproduces two essential aspects of the elastic differential cross section,  the well known precipitous descent in the forward direction and a sharp  `dip' structure. To include a complete description of data sets  near $-t=0$, we
  correct the original parametrization. We examine  two possibilities, the presence of the  two-pion threshold singularity or  a multiplicative factor
 reflecting  the proton form factor.  We find 
good descriptions of LHC7 and ISR data in either case. The form factor model allows simple predictions for higher energies through  asymptotic theorems and   asymptotic sum rules in impact parameter space. We present  predictions for this model at higher LHC energies, which can be used to test whether asymptotia is reached. The black disk limit in this model is seen to be reached only for $\sqrt{s}\sim 10^6\ TeV$.
\end{abstract}

\keywords{Elastic cross section, Asymptotia}
\pacs{13.75.Cs, 13.85.-t}

\maketitle
\tableofcontents

\section{Introduction}
The total $pp$ cross-section and the elastic differential cross-section offer a unique opportunity to study confinement and the transition to perturbative QCD, as they are influenced by large and small distances.

 We now have data for the total and the elastic differential cross section  from LHC running at $\sqrt{s}=7\ TeV$ (LHC7) \cite{Antchev:2013gaa}.
Data from   LHC running at $\sqrt{s}=8 \ TeV$ (LHC8), { soon to be available},    and $14\ TeV $ (LHC14) may be our last chance to explain $pp$ scattering  in fundamental terms. A tool to help in this endeavor is  a good phenomenological understanding of their energy behavior, without the bias imposed by models. To present one such phenomenological description  is the aim of this paper.

In what follows we shall propose an empirical description of the differential elastic $pp$ cross-section to be used at LCH8 and LHC14.  This description follows from the original proposal by Barger and Phillips  (BP) \cite{Phillips:1974vt}, who described ISR data with a 5 parameters fit, i.e. writing the scattering amplitude as
\begin{equation}
{\cal A}(s,t)=i[\sqrt{A(s)}e^{B(s)t/2}+ e^{i\phi(s)} \sqrt{C(s)}e^{D(s)t/2}].
\label{eq:bp0}
\end{equation}
In  \cite{Grau:2012wy}, we had applied this parametrization to preliminary TOTEM results at LHC7 elastic differential cross-section data \cite{Antchev:2011zz}. In this paper we refine that analysis %Physics Letter B 2012,
 presenting an improved description of published data \cite{Antchev:2013gaa}, which includes 
   the very small $-t$ value,  i.e. parametrizing both the total and the elastic cross-section within a few percent of the present LHC7 data.  
   We also  propose its extension to higher energies, providing a parametrization  obeying  asymptotic theorems \cite{Froissart:1961ux,Martin:1962rt} and apply it to study the black disk limit approach.

  \section{The  Barger and Phillips  model and  LHC7 data }
  The parametrization, given in  Eq. (\ref{eq:bp0}), corresponds to a complex amplitude, which is composed of two terms, and a relative phase $\phi $, which  was  found phenomenologically to be $\sim 2.8 \ rad $ { at LHC7}, and can be interpreted as corresponding to contributions from opposite parities, $C=\pm1$. The publication of the actual data by the part of the TOTEM collaboration \cite{Antchev:2013gaa}
  %Antchev:2012prep} 
  requires an update and also a revision of the analysis  we performed to preliminary TOTEM data in \cite{Grau:2012wy},\footnote{We note here that an error had occurred in \cite{Grau:2012wy} when  fitting   ISR data at $\sqrt{s}=53\ GeV$, with inclusion of 63 GeV data in the fit. Conclusions,  however, remain  unchanged.}.
Applying Eq. (\ref{eq:bp0}) to the published data and using the same parameters  of \cite{Grau:2012wy}, we find that, when both statistical and systematic errors are included in the fit, the  description is still acceptable with $\chi^{2}/DOF \sim 2.6$.
 However, when the analysis is performed with only statistical errors,  the $\chi^2$ for the entire range becomes unacceptably large.  { In particular, the parameterization of Eq.(\ref{eq:bp0}) reproduces poorly the measured value of the total cross-section at LHC7.}
 The problem therefore seems to lie with  the optical point. 
To pinpoint the origin of the problem, we have fitted the  now released  data  \cite{Antchev:2013gaa}
%Antchev:2012prep}
   implementing different cuts of $t_{min}$ for which the BP  model provides a suitable description.  Specifically, any result with $\chi^{2}/DOF\lesssim 3$ is considered acceptable. In Table \ref{t:lhcbp0} we display a grid of possible cuts and the respective $\chi^{2}/DOF$ values.   
We also calculate  the corresponding values obtained for the differential cross-section at the optical point and the total cross-section. We find that, when the fit with the BP amplitude is able to reproduce the optical point,  the statistical description is not very good. On the other hand, the fit becomes quite good for 0.2 GeV$^2<|t_{min}|<0.3$ GeV$^2$, even though the total cross-section obtained in these cases is too low.

\begin{figure}[H]
\centering
\includegraphics[width=12cm,height=8.0cm]{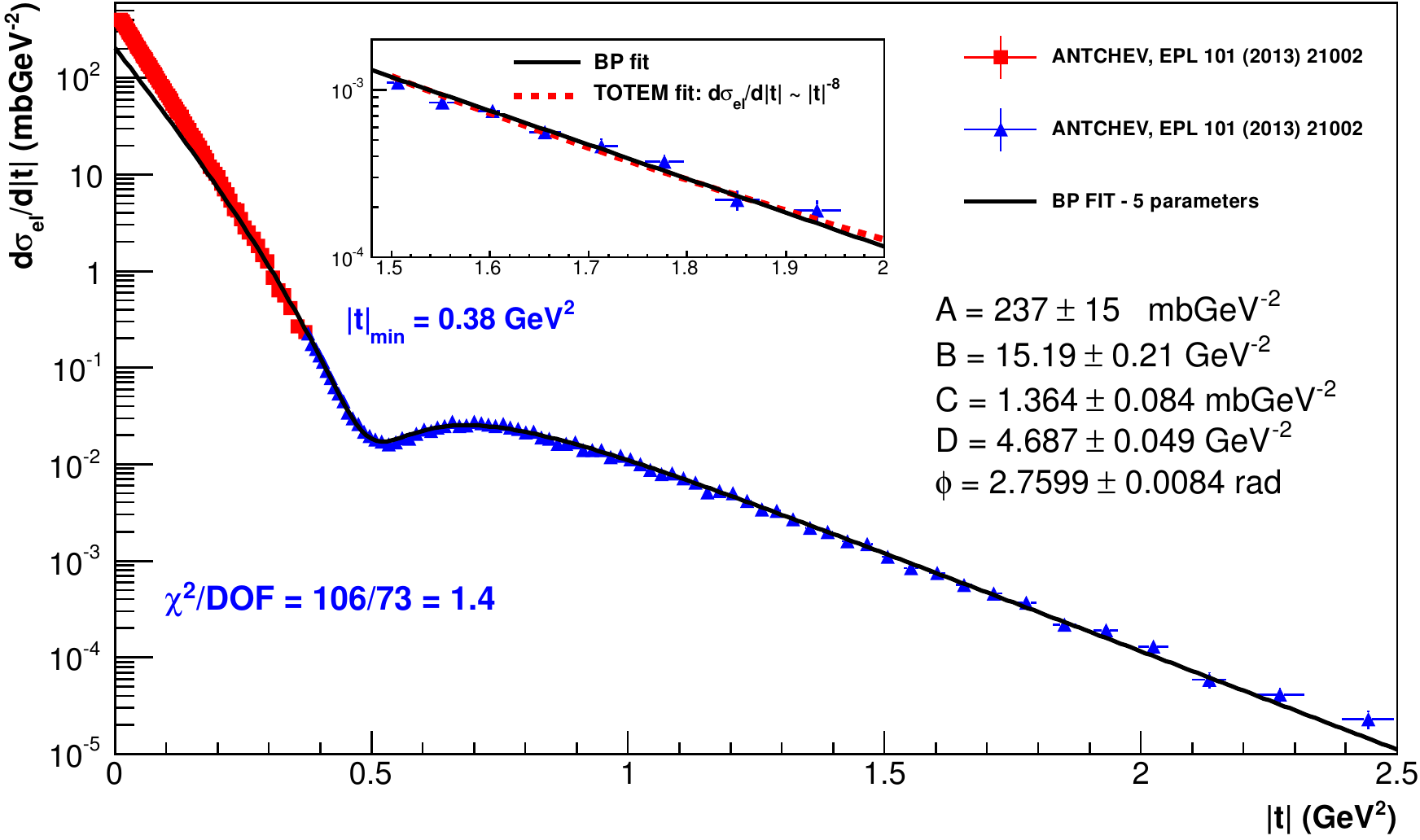} 
\caption{Fit to the differential $pp$ cross-section at 7.0 TeV  \cite{Antchev:2013gaa} 
with the BP parametrization of Eq. (\ref{eq:bp0}) in the range $0.38 \leq |t| \leq 2.4$ GeV$^{2}$ with $\chi^2/d.o.f$ in this interval. \textit{Inset}: the power law fit $|t|^{-n}$, with $n\approx 8$, compared to the exponential fit in the range $1.5 \leq |t| \leq 2.0$ GeV$^{2}$.}
\label{g:lhcbp2}
\end{figure}

\begin{table}[H]
\caption{Statistical results of fits with simple BP model of Eq. (\ref{eq:bp0}), { with  $\chi^2$ calculated  for the range  $-t>-t_{min}$ and resulting} values for  the optical point and the total cross section.}
\centering
\begin{tabular}{c|c|c||c|c}
\hline\hline $-t_{min}$ (GeV$^{2}$) & $DOF$ & $\chi^{2}/DOF$ & $d\sigma_{el}/dt\left|_{t=0} \right.$ (mbGeV$^{-2}$) & $\sigma_{tot}$ (mb) \\ 
\hline 0.01 & 156 & 9.40 & 490.2 & 97.9\\ 
\hline 0.10 & 118 & 6.33 & 422.8 & 90.9 \\ 
\hline 0.20 & 94 & 2.66 & 282.0 & 74.2\\ 
\hline 0.30 & 80 & 1.62 & 181.8 & 59.6 \\
\hline 0.40 & 70 & 1.41 & 212.1 & 64.4 \\ 
\hline\hline 
\end{tabular} 
\label{t:lhcbp0}
\end{table}
From Table \ref{t:lhcbp0} we conclude that, past the very small $-t<0.2 \ GeV^2$ values,  the  parametrization of Eq. (\ref{eq:bp0}) is suitable to describe two  essential features of 
the differential elastic cross section at high energies, namely the dip structure and the larger $|t|$ region, as 
one can also see from  Fig. \ref{g:lhcbp2}. Notice that  the exponential fit in the range $|t|>1.0$  GeV$^{2}$ can be taken to be as good as the power law fit $|t|^{-n}$ presented by the TOTEM Collaboration in \cite{Antchev:2011zz}, as  shown in  the inset of Fig. \ref{g:lhcbp2}.
We notice at this point that, for the BP  model of Eq. (\ref{eq:bp0}) to give    a good global description from the optical point to past the dip, the very small $|t|$ behavior must receive a correction,  while, at the same time, the region past $-t=0.2\ GeV^2$ should still be described { through  two exponentials (and the phase)}. 
Namely, since the BP model gives a very good description of LHC7 data except that in the forward region, there is  no phenomenological reason   to modify it neither around the dip nor in the tail. We thus propose to ameliorate the very small $-t$ behavior  by modifying only the first term in Eq. (\ref{eq:bp0}) with a factor $G(s,t)$ such that  $G(s,0)=1$, and suggest to parametrize existing and future $pp$ data with the amplitude:
\begin{equation}
\mathcal{A}(s,t)=i[G(s,t)\sqrt{A(s)}e^{B(s)t/2}+e^{i\phi(s)} \sqrt{C(s)}e^{D(s)t/2}]. \label{eq:mbp}
\end{equation}
We have examined   two possibilities:
\begin{itemize}
\item a  factor $G(s,t)=exp[-\gamma(s)(\sqrt{4\mu^2-t}-2\mu)]$ reflecting the presence of the nearest $t$-channel singularity, i.e. the two pion threshold  \cite{CohenTannoudji:1972gd,Anselm:1972ir} discussed in \cite{Khoze:2000wk} and \cite{Fiore:2008tp}, labeling this possibility as the  $mBP1$ model,
\item a factor $G(s,t)=F^2_P=1/(1-t/t_0)^4$ modeled after the proton form factor, which describes phenomenologically the probability that the proton breaks up as the squared momentum transfer increases, labeling this as $mBP2$ model. 
\end{itemize}
The   $mBP1$ model, and other possible modifications of the BP models are discussed in detail  in Appendices  \ref{app:2pion} and \ref{app:ffBP} of this paper.
In  the appendices 
% \ref{app:impact}
  we shall 
also present  further details,  such as analytic  expressions for the  elastic cross-section and, {
 in Appendix \ref{app:impact}, } the impact parameter profile functions for both $mBP1$ and $mBP2$ models.

\section{The proton form factor modification: {\it mBP2}\label{sec:mbp2}}
\begin{figure}[H]
\begin{center}
\includegraphics*[width=8.5cm,height=7.5cm]{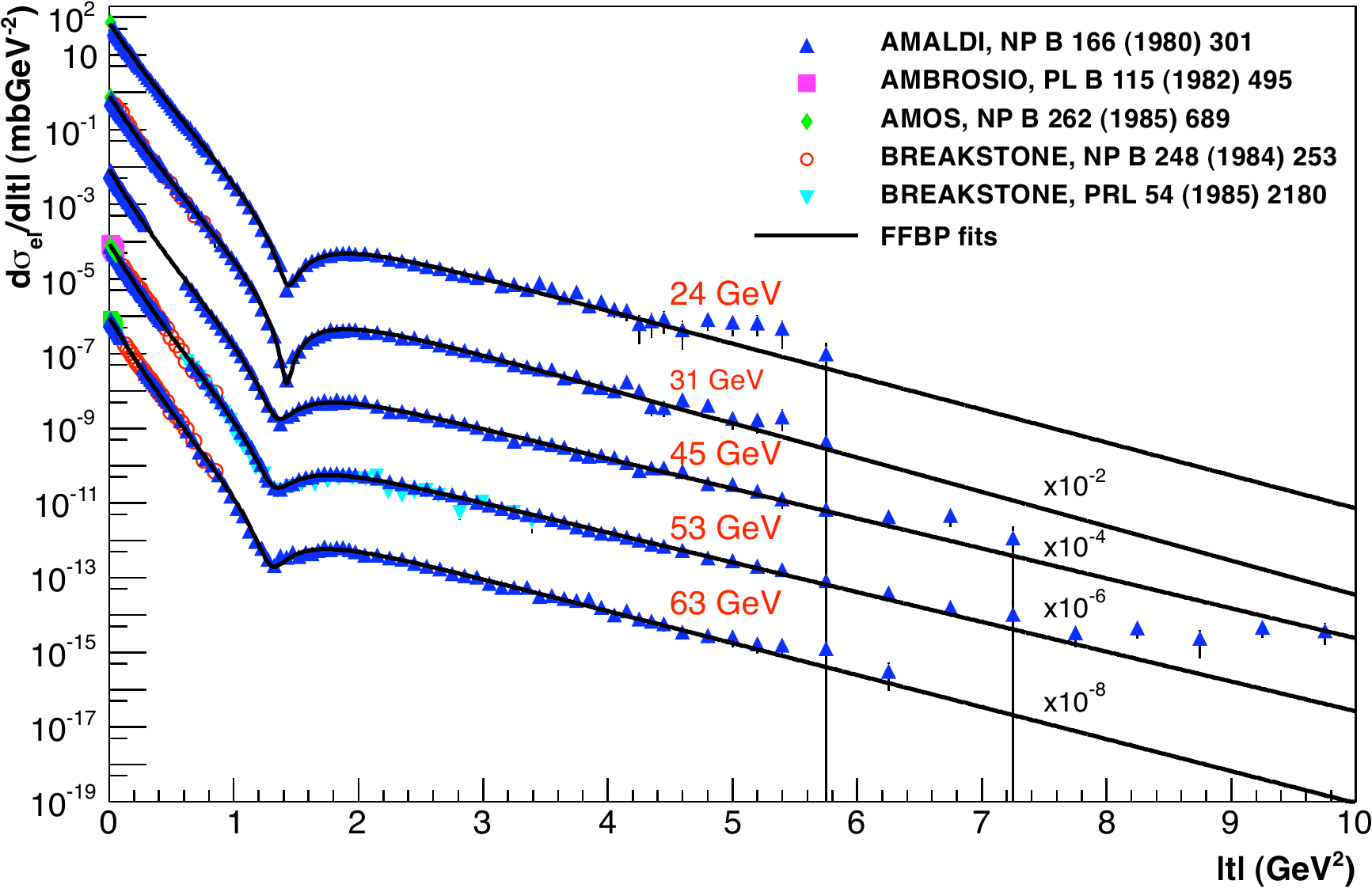}\vspace*{0.2cm}
\includegraphics*[width=8.5cm,height=7.5cm]{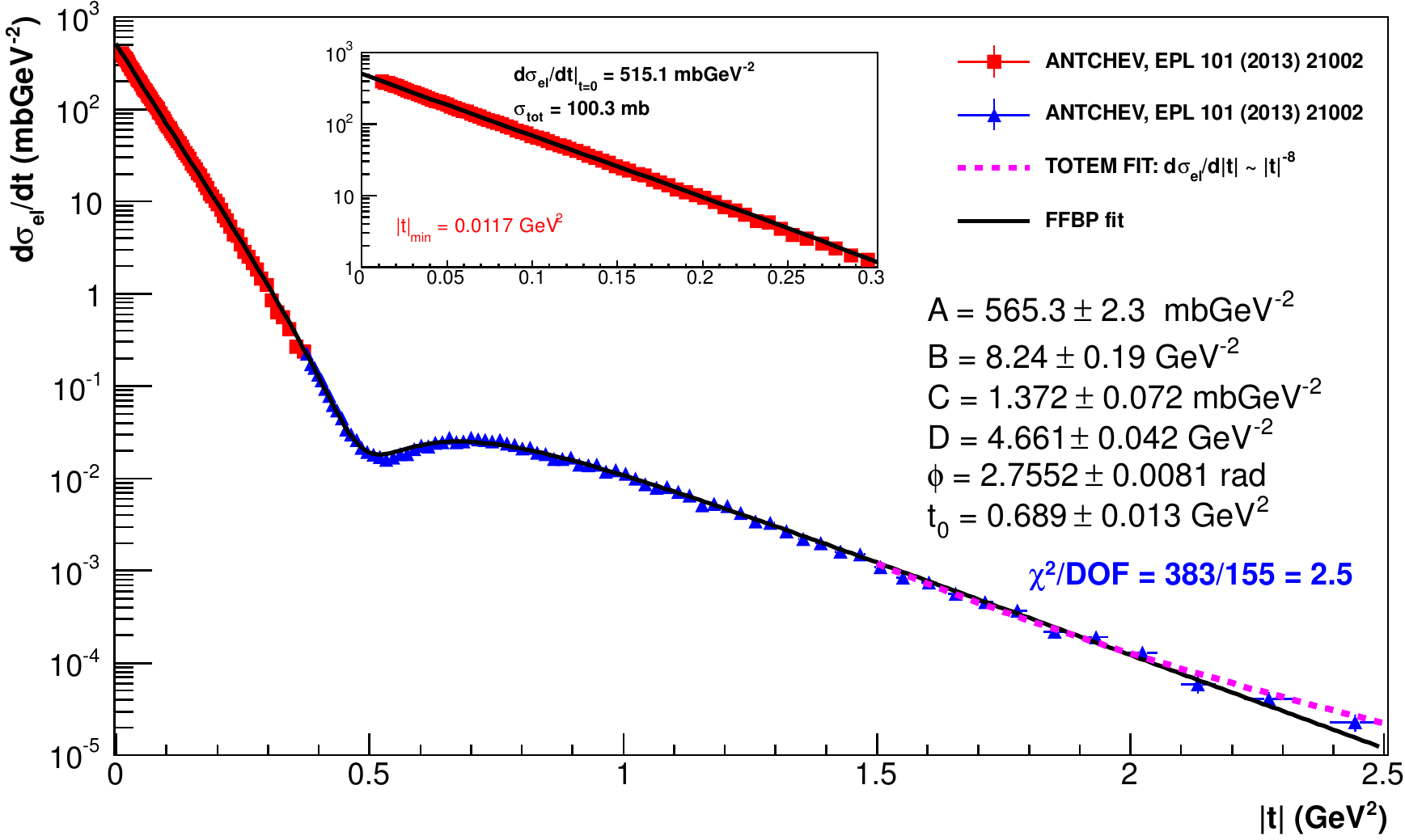}
\caption{Fits to the ISR and LHC7 data sets with model $mBP2$,  and $t_0$ a free parameter. Data sets as described in the text. }
\label{g:lhcbp5}
\end{center}
\end{figure}

 In this section, and the ones to follow, we   present here a    modification of the BP model at very small $t$-values obtained  through the proton form factor. As  viable parametrization of LHC data, 
 % a modified BP model, in which only the first term is changed,  namely  
 we analyze the physics content of the following  model for the elastic scattering amplitude: 
\begin{equation}
\mathcal{A}(s,t)=i[F^2_P(t)\sqrt{A(s)}e^{B(s)t/2}+e^{i\phi(s)} \sqrt{C(s)}e^{D(s)t/2}] \label{eq:bp7}
\end{equation}
with $A,\ B,\ C,\ D,\phi$ and $t_0$ as free parameters.
We display our results with this model (henceforth called $mBP2$) in Fig. \ref{g:lhcbp5}. { ISR data sets used in the fits  comprise the data collection by Amaldi and Schubert \cite{Amaldi:1979kd} and all experimental information available from 1980 onwards \cite{Amos:1985wx,Breakstone:1984te,Breakstone:1985pe,Ambrosio:1982zj}.
 LHC7 data are from \cite{Antchev:2013gaa}.} 
 We summarize the results of the fit in Table \ref{t:lhcbp2}, with the last two rows corresponding to the fit to LHC7 data obtained by using for $t_0$ the results from the fit or keeping $t_0$ fixed at $0.71\ GeV^2$. 
 Plots and fits results for the model with the two pion threshold correcting the $t\sim 0$ behavior can be found in Appendix \ref{app:2pion}.
\begin{table}[H]
\vspace*{-0.5cm}
\caption{
First six  rows give  values of free fit parameters $A, B, C, D, t_{0}$ and $\phi$ for the model $mBP2$ at each energy analyzed. In the last row, the scale parameter $t_0$ is kept fixed. $A$ and $C$ are expressed in units mbGeV$^{-2}$, $B$ and $D$ in units GeV$^{-2}$, $t_{0}$ in units GeV$^{2}$ and $\phi$,  in radians.} 
\vspace{0.2cm}
\centering
\renewcommand{\arraystretch}{1.5}
{\scriptsize
\begin{tabular}{|c|c|c|c|c|c|c||c|c|}
\hline \hline 
$\sqrt{s}$ (GeV) & $A$ & $B$ & $C (\times 10^{-3})$  & $D$  & $t_{0}$  & $\phi$ & $DOF$ &$\frac{\chi^{2}}{\rm DOF}$\\ 
\hline 24 &  74.8 $\!\pm\!$ 0.8 & 4.0 $\!\pm\!$ 0.1 &  4.8 $\!\pm\!$ 0.7  &  2.03 $\!\pm\!$ 0.06
  & 1.06 $\!\pm\!$ 0.03  & 3.31 $\!\pm\!$ 0.01  & 128 & 1.2 \\  
\hline 31 & 83.7 $\!\pm\!$ 0.2   &  3.90 $\!\pm\!$ 0.07  &  5.4 $\!\pm\!$ 0.5  & 2.12 $\!\pm\!$ 0.04 
&  0.99 $\!\pm\!$ 0.01  & 3.06 $\!\pm\!$ 0.01  & 200 & 1.6 \\ 
\hline 45 & 89.6 $\!\pm\!$ 0.2   &  4.27 $\!\pm\!$ 0.05  & 2.4 $\!\pm\!$ 0.2  &  1.84 $\!\pm\!$ 0.02
& 0.912 $\!\pm\!$ 0.009   & 2.83 $\!\pm\!$ 0.01 & 201 & 3.7 \\  
\hline 53 & 93.0  $\!\pm\!$ 0.1   & 4.51 $\!\pm\!$ 0.05  & 2.5 $\!\pm\!$ 0.1   & 1.84 $\!\pm\!$ 0.01   & 0.947 $\!\pm\!$ 0.008  & 2.79 $\!\pm\!$ 0.01   & 313 & 4.7 \\ 
\hline 63 & 97.4  $\!\pm\!$ 0.2   & 4.3 $\!\pm\!$ 0.1   & 3.5 $\!\pm\!$ 0.4   & 1.97 $\!\pm\!$ 0.04  & 0.90 $\!\pm\!$ 0.01 &  2.86 $\!\pm\!$ 0.06   & 159 & 2.1 \\
\hline 7000 & 565 $\!\pm\!$ 2  & 8.2 $\!\pm\!$ 0.2 &  1370 $\!\pm\!$ 70   & 4.66 $\!\pm\!$ 0.04 & 0.69 $\!\pm\!$ 0.01 & 2.755 $\!\pm\!$ 0.008 & 155 & 2.5\\
\hline 7000 & 562 $\!\pm\!$ 1  & 8.54 $\!\pm\!$ 0.03 &  1280 $\!\pm\!$ 34   & 4.61 $\!\pm\!$ 0.03 & 0.71 (fixed)  & 2.744 $\!\pm\!$ 0.004 & 156 & 2.5\\
\hline \hline
\end{tabular}}
\label{t:lhcbp2}
\end{table} 

We notice that the value of the parameter $t_0$  is  larger at ISR energies than at LHC7, where  its value is consistent with $F_P(t)$ being  the  EM form factor, i.e. $t_0\sim 0.71 \ GeV^2$. { In fact, fits to the LHC7 data with this value give  a $\chi^2/{\rm DOF}=2.5$ just as in the case of the free fit.  This difference between ISR and LHC probably corresponds to low energy contributions to this parameter. We  make the ansatz that when asymptotic energies  are reached,   $t_0$  correspond to   its EM form factor value. 
At non-asymptotic energies,   the parameter $t_0$ can be parametrized as shown in Fig. ~\ref{fig:t0}. In this figure,  elsewherelse in this paper and unless otherwise stated, the squared energies $s$ in  $\ln s$ are    in units of $GeV^2$. 
\begin{figure}[H]
\centering
\includegraphics[width=8 cm,height=6cm]{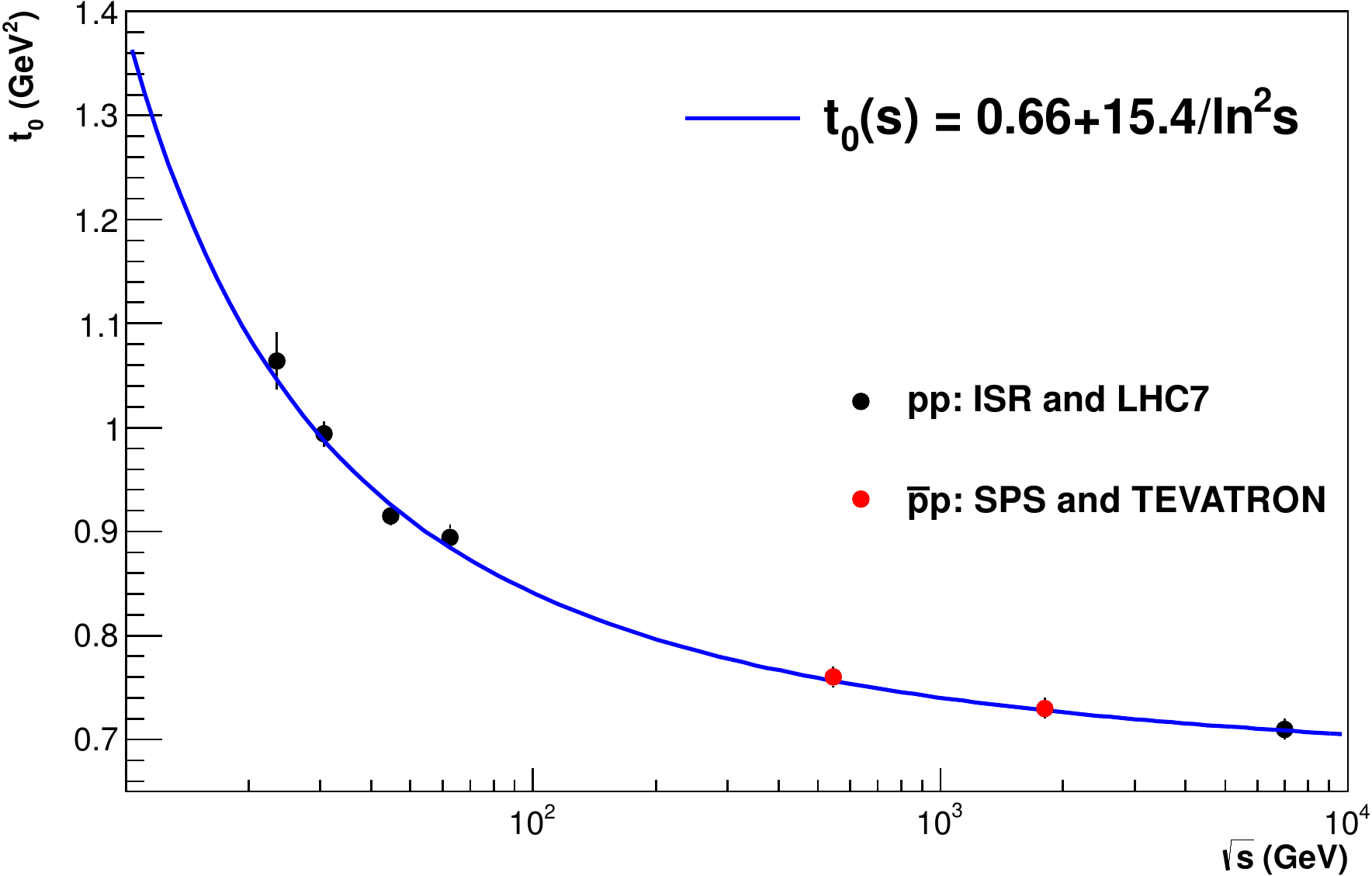}
\caption{Fit to the energy dependence of the scale parameter $t_0$. The black points correspond to results from Table \ref{t:lhcbp2} for $pp$. The red dots indicate the value taken by this parameterization for $t_0$ at energies corresponding to $S{\bar p}p$ and the TeVatron, where the process is $\pbarp$.}
\label{fig:t0}
\end{figure}

 From the $mBP2$ model the analytical expression for the elastic cross section follows:
\begin{equation}
\sigma_{el}(s)=At_0e^{Bt_0}E_8(Bt_0)+\frac{C}{D}+2(\sqrt{AC}\cos \phi)t_0
e^{(B+D)t_0/2}E_4(\frac{(B+D)t_0}{2})
\end{equation}
with  $E_{n}(x)=\int_1^\infty dy e^{-xy}/y^n$. In Table \ref{t:lhcbp3} we present the values of the total and elastic cross sections as obtained from both models, $mBP1$ and $mBP2$,  together with the optical point for LHC7 and ISR energies. Parameter values for the $mBP1$ model can be found in Appendix \ref{app:2pion}.

 %%%%%%%%%%%%%%%%%%%%%%%%%%%%%%%%%%%%%%Section4%%%%%%%%%%%%%%%%%%%%%%%%%%%%%%%%%%%%%%%%%%%%%%%%

\begin{table}[H]
\caption{Cross sections and the optical point following from models $mBP1$ and $mBP2$.}
\centering
\vspace*{0.2cm}
\begin{tabular}{c|c|c|c|c}
\hline\hline Model & $\sqrt{s}$ (GeV) & $\sigma_{tot}$ (mb) & $\sigma_{el}$ (mb) & $d\sigma_{el}/dt\left|_{t=0} \right.$ (mbGeV$^{-2}$)  \\ 
\hline \multirow{6}{*}{$mBP1$} & 24 & 40.0 & 6.80 & 82.0 \\ 
 & 31 & 40.6 & 7.15 & 84.3 \\ 
 & 45 & 42.1 & 7.14 & 90.9 \\ 
 & 53 & 42.9 & 7.43 & 94.0 \\ 
 & 63 & 43.7 & 7.68 & 97.8 \\ 
 & 7000 & 101 & 25.6 & 524 \\ 
 \hline
 \multirow{6}{*}{$mBP2$} & 24 & 37.9 & 6.65 & 73.6 \\ 
 & 31 & 40.1 & 7.20 & 82.4 \\ 
 & 45 & 41.6 & 7.13 & 88.7 \\ 
 & 53 &  42.4 & 7.42 & 92.1  \\ 
 & 63 & 43.3 & 7.60 & 96.3 \\ 
 & 7000 & 100 & 25.5 & 515 \\ 
\hline \hline
\end{tabular} 
\label{t:lhcbp3}
\end{table}

We shall now apply to  the modified amplitude 
 the asymptotic sum rules presented in  our previous analysis \cite{Grau:2012wy}.  
 The sum rules correspond to the ansatz of total absorption in $b$-space. { Namely, $SR_1\equiv \Im m {\tilde {\cal A}}(s,b=0)=1$ and $SR_0\equiv \Re e {\tilde {\cal A}}(s,b=0)=0$ at asymptotic energies, where ${\tilde {\cal A}}(b,s)$ is the Fourier transform of the scattering amplitude in $b$-space. 
 For the $mBP1$ model, the analytical expressions for the sum rules for imaginary and real part  of the amplitude are presented in  Appendix \ref{app:2pion}.  As discussed in \cite{Grau:2012wy},  for the satisfaction of the first sum rule, $SR_0=0$, it is necessary to introduce a real part for the first term, the one dominant at small $-t$, for which $C=+1$.  Let us denote with ${\hat \rho}(s) $ the  contribution to the ratio of the real to the imaginary part of the first term.}
 For $mBP2$ the sum rules give the following results
\begin{eqnarray}
SR_1 = \frac{1}{2\sqrt{\pi}} \int_o^\infty dT [ \sqrt{\frac{A}{1 + {\hat \rho}^2}}  \frac{e^{- B T/2}}{[1 + (T/t_o)]^4} -  \sqrt{C} e^{- DT/2} |cos \phi|]=\label{3.12}\\
= \frac{1}{\sqrt{\pi}} [ - \frac{\sqrt{C}}{D} |\cos \phi| + \sqrt{\frac{A}{1 + {\hat \rho}^2}}\frac{t_0}{2} e^{Bt_0/2} E_4(Bt_0/2)] \\
SR_0= \frac{1}{\sqrt{\pi}} [ - \frac{\sqrt{C}}{D} \sin \phi +  \sqrt{\frac{A}{1 + {\hat \rho}^2}}  \hat{\rho} \frac{t_0}{2} e^{Bt_0/2} E_4(Bt_0/2)]; \; 
\label{3.2}
\end{eqnarray}
{ Using the tight bound}
\begin{equation}
\label{3.5}
 \frac{1}{x + n}   < [e^x E_n(x)] < \frac{1}{x + n - 1};\ n = 1, 2,...
\end{equation}
{\ we have} 
\begin{equation}
\label{3.7}
[\frac{1}{ B + 8/t_0}] < \frac{t_0}{2}e^{Bt_0/2}E_4(B t_0/2) < [\frac{1}{ B + 6/t_0}].\label{eq:bav}
\end{equation}
{ Hence, a simple analytical result for the sum rules can be obtained  in the modified  $mBP2$ model by taking  the mean value $7$ in the denominator,  so that }
\begin{eqnarray}
SR_1 = \frac{1}{\sqrt{\pi}} [ - \frac{\sqrt{C}}{D} |\cos \phi| + \frac{ \sqrt{\frac{A}{1 + {\hat \rho}^2}} }{\hat{B}}]; \ \hat{B} = B + \frac{7}{t_0} \label{3.8}\\
SR_0= \frac{1}{\sqrt{\pi}} [ - \frac{\sqrt{C}}{D} \sin \phi + \frac{ \sqrt{\frac{A}{1 + {\hat \rho}^2}} }{\hat{B}} \hat{\rho}]; \ \hat{B} = B + \frac{7}{t_0}. \label{3.9}
\end{eqnarray}
{ Asymptotically, we expect the following:}
\begin{equation}
\label{3.10}
SR_1 \to\ 1- ;\ SR_0 \to\ 0+.
\end{equation}
{In order to estimate the values taken by $SR_0$ and $SR_1$ and check whether total absorption has been taking place, an estimate for ${\hat \rho}$ is needed. To this aim we use  the  soft $k_t$ resummation model of Ref. \cite{Godbole:2004kx} where the leading term of the cross-section is driven by QCD mini-jets. 
In this model the asymptotic behavior of the total cross-section is obtained as $\sigma_{total}\sim (\ln s)^{1/p}$  \cite{Grau:2009qx}, where the parameter $p$ controls the large $b-$ behavior of the impact parameter distribution and obeys the constraint $1/2<p<1$. Asymptotically then,  ${\hat \rho}(s)=\pi/2p\ln s$. }

}We show  in Table \ref{t:lhcbp5} the numerical results for $SR_1$ and $SR_0$ 
for both models, $mBP1$, where the small $-t$ modification is obtained through a term reflecting the two pion loop  singularity, and $mBP2$, where the form factor is dominating the $t\simeq 0$ behavior. 
\begin{table}[H]
\caption{Sum rules for modified BP models at ISR23, ISR53 and LHC7.}
\centering
\vspace*{0.2cm}
\begin{tabular}{c|c|c|c|c}
\hline\hline Model & $p$ & $\sqrt{s}$ (GeV) & $SR_{1}$ & $SR_{0}$ \\ 
\hline
\multirow{4}{*}{$mBP1$} & $-$ & 24 & 0.721 & 0.011\\
& $-$ & 53 & 0.722 & 0.049\\
& 0.66 & 7000 & 0.953 & 0.067\\
& 0.77 & 7000 & 0.956 & 0.046\\ \hline
\multirow{4}{*}{$mBP2$} & $-$ & 24 & 0.719 & 0.021 \\
& $-$ &  53 &  0.717 & 0.049 \\
& 0.66 &  7000 &  0.950 & 0.070 \\
& 0.77 & 7000 & 0.953 & 0.048 \\
\hline\hline
\end{tabular} 
\label{t:lhcbp5}
\end{table}
The table indicates that the modified models ameliorate the satisfaction of the Sum Rules with respect to the simpler BP parametrization, but the asymptotic value $SR_1=1$ is not yet reached.

{ We also notice that in the BP  model (and in its modified versions as well),  the parameter $\rho(s)$,  real to imaginary part in the forward direction, is given by}
\begin{equation}
\label{4.1}
\rho(s) = \frac{\hat{\rho} - \sqrt{(\frac{C}{A})}\ sin \phi}{1 -  \sqrt{(\frac{C}{A})} |cos \phi|} \longrightarrow { \hat\rho} + \sqrt{(\frac{C}{A})} [{\hat \rho} \ |\cos \phi| - \sin \phi ]. \label{eq:rhovsc}
\end{equation}
  for $\sqrt{C/A}<< 1$.
 We shall make use of Eq.~(\ref{eq:rhovsc}) when discussing asymptotic predictions.

\begin{figure}[H]
\includegraphics[width=8cm,height=6cm]{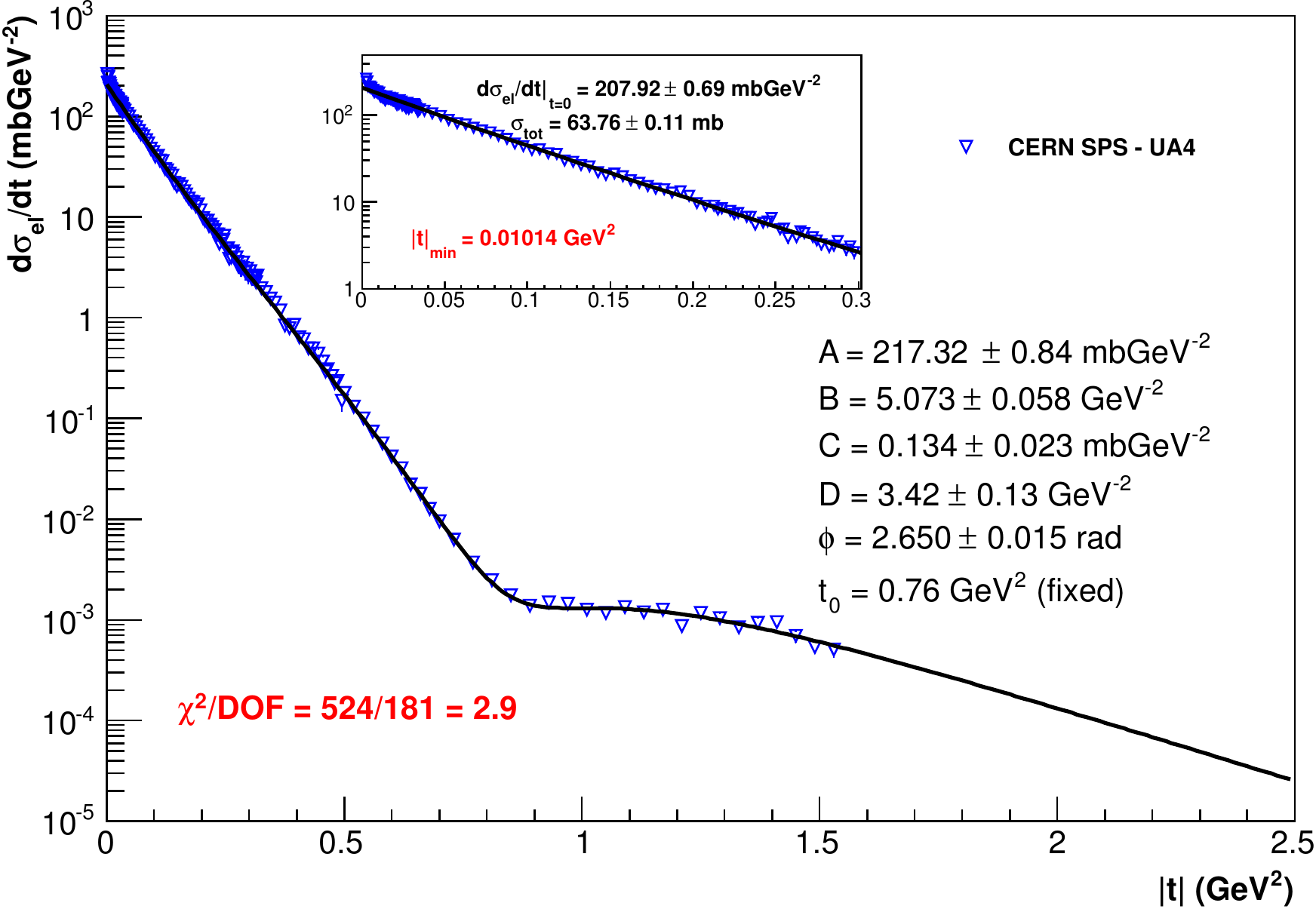}
\includegraphics[width=8cm,height=6cm]{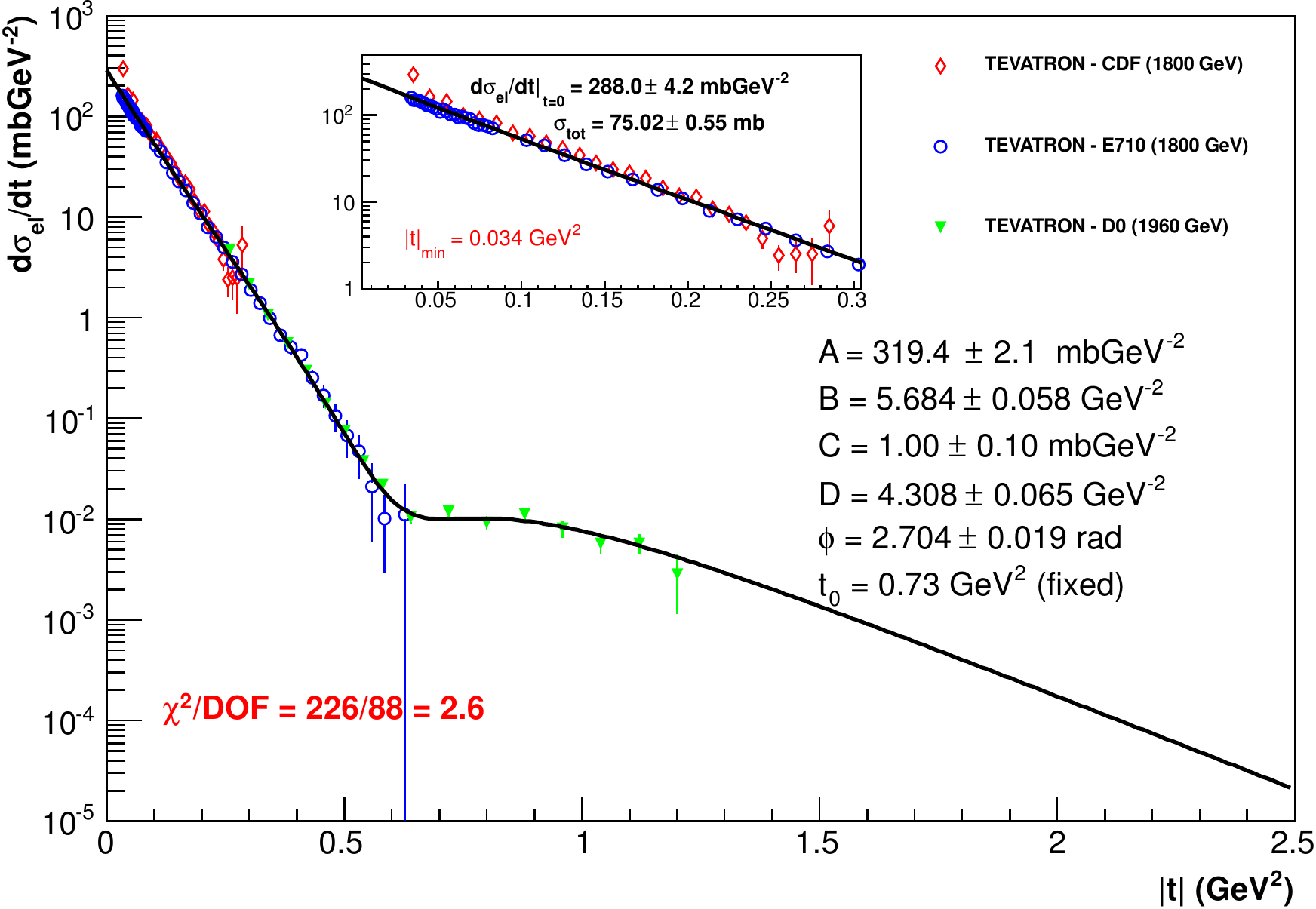}
\caption{The $mBP2$ model applied to $\pbarp$ data. Value for parameters thus obtained are shown in the plots.}
\label{fig:pbarp}
\end{figure}

 Before leaving this section, we present the results one obtains when the model $mBP2$ is applied to elastic $\pbarp$ data. Following the previous comments, the scale   $t _0$ has been fixed from the parametrization shown in Fig. \ref{fig:t0}. We show the results of the fits to $\pbarp$ data  in 
 Fig. \ref{fig:pbarp}.
{\ We note the absence of a distinctive dip in $\pbarp$, but also its  faint appearance as the energy increases.}

\subsection{Slope parameter in the modified models}

The introduction of a \textit{general} factor, $G(s,t)$, given either as  in Eq.~(\ref{eq:bp2}) or by the square of the proton form factor, 
 leads to a change of curvature in the local slope. This behavior should be expected since the new model is influenced by $G(s,t)$ as follows:

\begin{eqnarray}
B_{eff}(s,t) &=& \left( \frac{d\sigma_{el}}{dt} \right)^{-1} \left[ ABe^{Bt}G^{2}(s,t)+2Ae^{Bt}G(s,t)\frac{d G(s,t)}{dt }+CDe^{Dt}\right.  \nonumber \\
&+& \left.\sqrt{AC}(B+D)G(s,t)e^{(B+D)t/2}\cos \phi + 2\sqrt{AC}e^{(B+D)t/2}\frac{d G(s,t)}{d t}\cos \phi \right]. \label{eq:bp18}
\end{eqnarray}

\par
\begin{figure}[H]
\begin{center}
\includegraphics*[width=8.5cm,height=7.5cm]{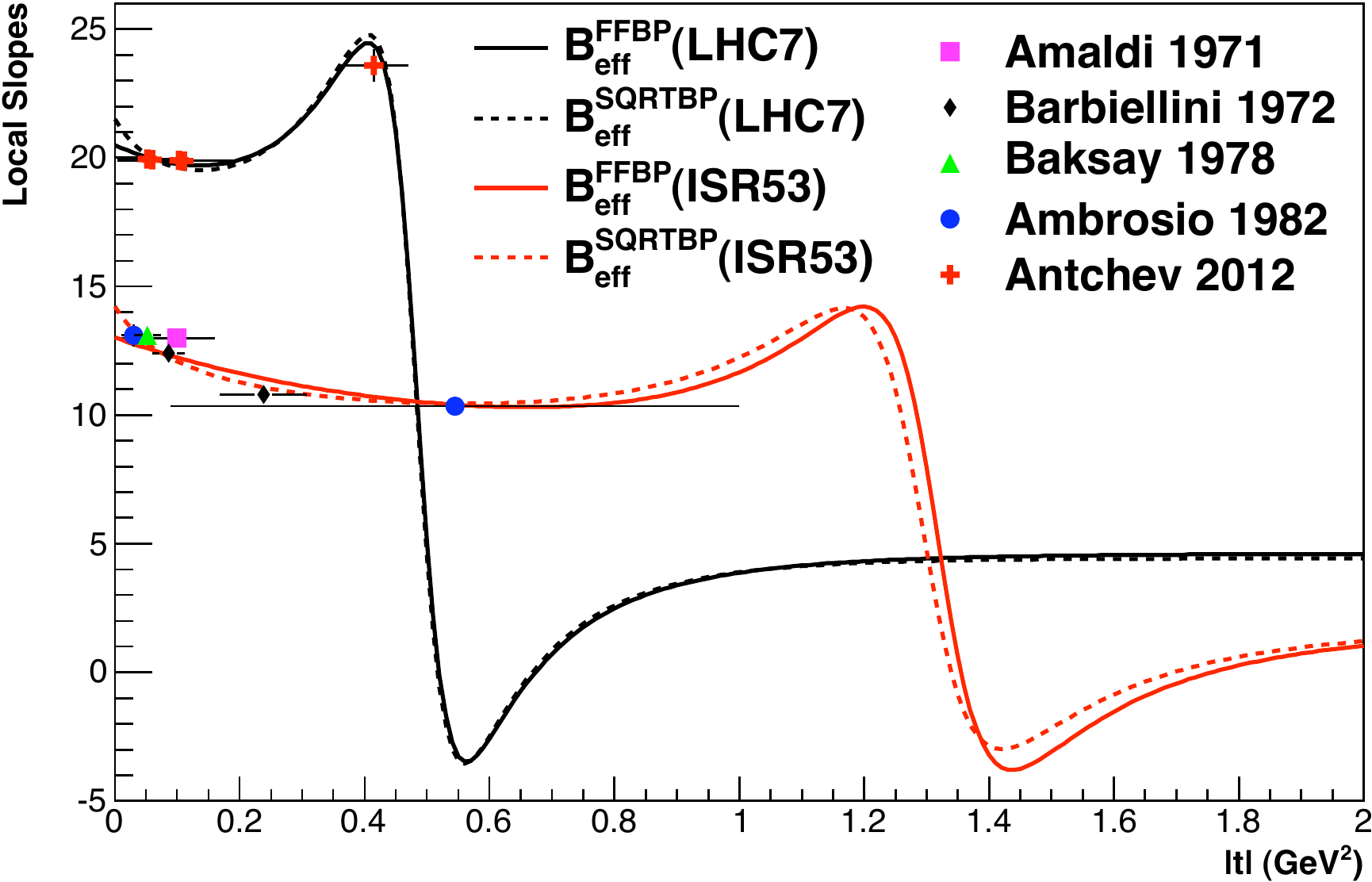}
\vspace{0.1cm}
\includegraphics*[width=8.5cm,height=7.5cm]{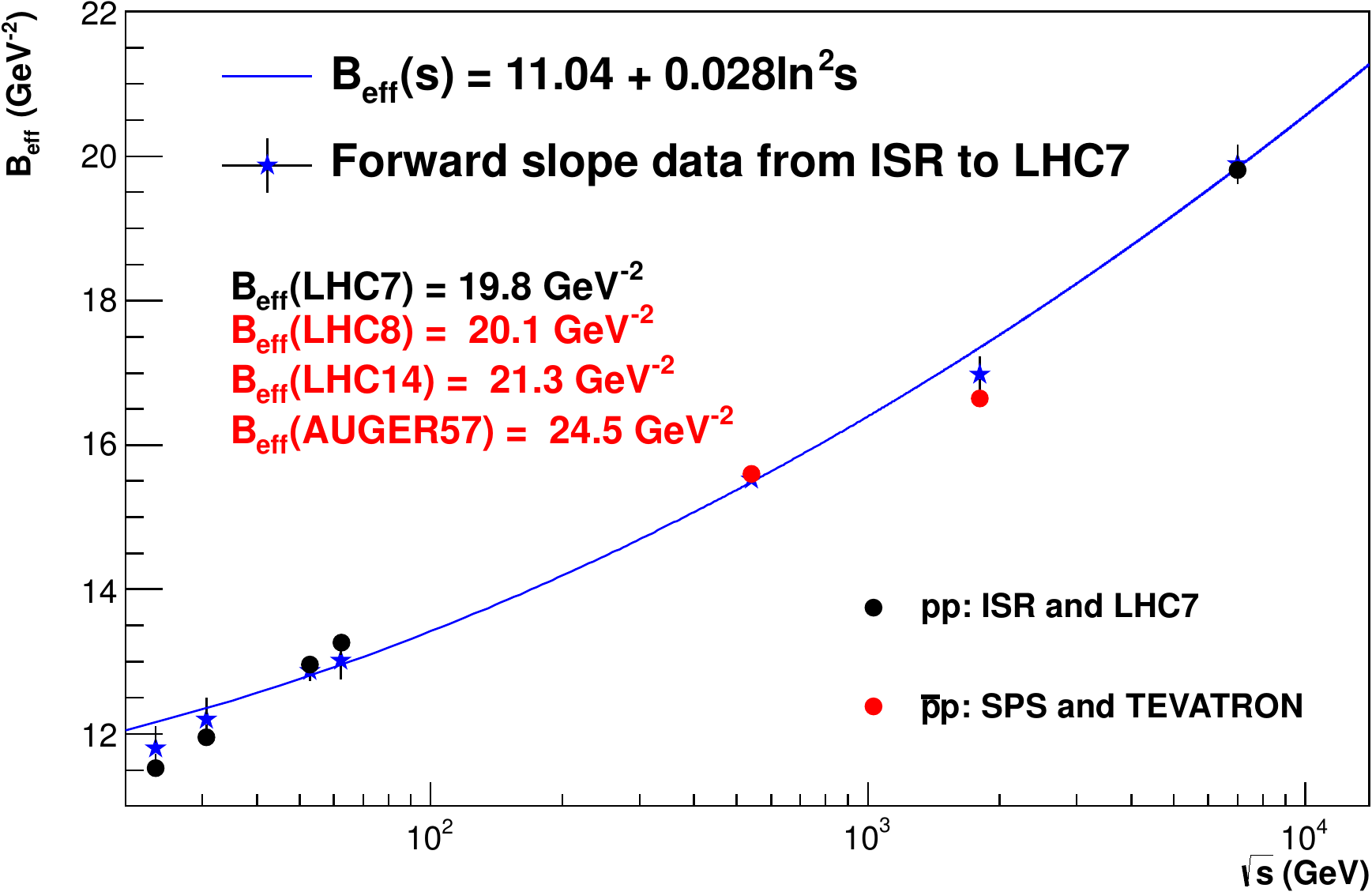}
%FIT_Beff_MARCH21-D}
%\end{center}
%\end{figure}
%\end{document}
\caption{At  left, local and forward slopes from the two-pion threshold model (dashes,  label $B_{eff}^{SQRTBP}$) and the form factor one (full line, label $B_{eff}^{FFBP}$) at ISR53 and LHC7. At  right, the effective  forward slope from both $pp$ and $\pbarp$ data (blue stars). Data are  compared with a fit in $(\ln s)^2$ and  from  the model  approximation  of Eq.~(\ref{eq:beffvsb}), using results from Table \ref{t:lhcbp2} for $pp$ (black dots). Red dots indicate the result for $\pbarp$, with   the scale $t_0$ obtained  from Fig. \ref{fig:t0}. }. 
\label{g:lhcbp6}
\end{center}
\end{figure}

\noindent
 In Fig.~\ref{g:lhcbp6} we display data for the effective { forward slope $B_{eff}(s)\equiv B_{eff}(s,t=0)$,} compared with  the local slope $B_{eff}(s,t)$, at ISR53 and LHC7, following from  the above-mentioned models. 
  We notice from this figure that the modification with the square root in the exponential, $mBP1$, appears to overrate near-forward slopes. In fact, the respective forward slopes exceed by some $10\%$ the measurements at ISR53 and LHC7.  This provides yet another   reason  to focus on the form factor modified model, $mBP2$.
%  \end{document}
{ We  also notice that, from a numerical point view,  while
in the original BP model $B(s) \simeq B_{eff}(s)$, in the case of $mBP2$ the following approximation holds:}
\begin{eqnarray}
B(s) \simeq B_{eff}(s) - \frac{8}{t_{0}} \label{eq:beffvsb}
\end{eqnarray}
 as one can easily check using Eq. (\ref{eq:bp18}).  For this model, we also show in  the right panel of  Fig. \ref{g:lhcbp6} the comparison between $B_{eff}(s)$ from experimental results from ISR to LHC.
 The  parametrization applied in Fig. ~\ref{g:lhcbp6} for $B_{eff}(s)$ is inspired  by the asymptotic theorems, and  consistent with  the result for the effective slope by Schegelski and Ryskin \cite{Schegelsky:2011aa}. Indeed, when fitting $B_{eff}(s)$ with an additional term with a linear $\ln s$ dependence, the respective coefficient is  consistent with zero. That leads to  the best fit shown in Fig. ~\ref{g:lhcbp6}, with  $B_{eff}(s) \sim (\ln s)^2$.
 
  \section{Asymptotic predictions of the empirical model $mBP2$}
  
The original BP model of Eq. (\ref{eq:bp0}) had purported to present   a "model independent analysis  of the structure in $pp$ scattering" \cite{Phillips:1974vt}. As such, and as recently pointed out by Uzhinsky  and Galoyan \cite{Uzhinsky:2012yx},
the BP parametrization does not, in itself, possess a predictive power. However,  its simplicity can be exploited  to make higher energy predictions.
In fact, the model  has the virtue of allowing a simple implementation of the asymptotic sum rules we presented in \cite{Grau:2012wy}, and thus to obtain the asymptotic behavior of the parameters  which can lead to {  this model} predictions for the elastic differential cross-section at LHC8 and LHC14. We shall now proceed to illustrate such an asymptotic, and  partly empirical,  realization of the $mBP2$ model.

This model  has 6 parameters, i.e. 2 amplitudes $\sqrt{A(s)}$ and $\sqrt{C(s)}$, two slopes $B(s)$ and $D(s)$, a phase $\phi$ and a scale $t_0$. The fits to ISR and LHC7 data suggest $t_0\rightarrow 0.71\ GeV^{2} $ at LHC energies, thus for asymptotic predictions we fix $t_0$ to acquire the value of the EM form factor of the proton, i.e. $t_0=0.71\ GeV^2$. As for the phase $\phi$, the same fits support the approximation $\phi \sim constant $  in energy. { In Regge models,  the phase would be  t-dependent,
and, in such case,  the phase, as used here in the empirical model, would represent  a  value averaged over  the range $
\Delta t $ of validity of this model.}

Having thus made the ansatz that both $t_0$ and $\phi$ are asymptotically constant, we  remain with 4 energy dependent parameters. As we shall  shortly discuss in detail, to comply with asymptotic theorems $\sqrt{A}$ and $B(s)$ should have the same asymptotic behavior, namely   at most like  $(\ln s )^2$. For the  slope of the second (non leading) term, an asymptotic normal Regge behavior would be the most appealing possibility. The amplitude of the second term is so far unconstrained. From the asymptotic sum rules, the amplitude $\sqrt{C(s)}$ can either have a constant or a logarithmic energy dependence.
We shall now see how  this behavior can be understood in more detail.

The satisfaction of the sum rules for elastic scattering at higher energies, namely, $SR_{1} \rightarrow 1$ and  $SR_{0} \rightarrow 0$, is suggested by our results, presented in Table \ref{t:lhcbp5}. Based on their saturation, we propose to make predictions  concerning the energy behavior of the parameters $A(s),\  B(s), \ C(s)$ and $D(s)$.
We begin with the simple BP model, which contains the asymptotics of the sum rules, since both $\gamma (s)$ and $t_0$ of the modified versions of the model are approximately  constant in energy. The asymptotic sum rules read:
\begin{eqnarray}
SR_0=\sqrt
{\frac{A(s)}{1+{\hat \rho}(s)^2}}
\frac{
{\hat \rho}(s)}
{
\sqrt{\pi}B(s)
}
-\frac{
\sqrt{C(s)}\sin\phi
}{
\sqrt{\pi}D(s)
}\rightarrow 0
\\
SR_1=\sqrt{
\frac{
A(s)
}
{1+{\hat \rho}(s)^2}
}\frac{1}{
\sqrt{\pi}B(s)}
+\frac{
\sqrt{C(s)}\cos\phi
}{
\sqrt{\pi}D(s)
}\rightarrow 1
\end{eqnarray}
Since 
$\phi$ is approximately constant throughout the range from ISR and beyond, and if ${\hat \rho}(s)\sim 1/\ln s$, 
one can then obtain the following asymptotic relationships between the parameters:
\begin{eqnarray}
\frac{\sqrt{A(s)}}{B(s)}\sim \frac{\sqrt{C(s)}}{D(s)}\ln s \label{eq:sr0}\\
\frac{\sqrt{A(s)}}{B(s)}\sim \frac{\sqrt{\pi} }{(1+\frac{\pi \cot \phi}{2p\ln s})}\sim constant \label{eq:sr1}
\end{eqnarray}

 We now start from the fact that
to leading order in $\ln s$, the parameter $A(s)\propto \sigma_{tot}^2$. 
The satisfaction of    asymptotic theorems \cite{Block:1984ru} suggests that asymptotically $\sigma_{total}\sim B(s)$, which is also in agreement  with Eq.~(\ref{eq:sr1}).

  We consider here a specific realization of the Froissart-Martin  bound \cite{Froissart:1961ux,Martin:1962rt}, namely 
the case of maximal {\it energy} saturation.  The more general case of $\sigma_{total }\sim (\ln s)^{1/p}$ with $1/2<p<1$ will be discussed elsewhere. Then, 

\begin{equation}
A(s)\sim (\ln s)^4, \ \ \ \ \ B(s)\sim (\ln s)^2,\ \ \ \ \ D(s)\sim \sqrt{C(s)} \ln s \label{eq:SRasympt}
\end{equation}
 The above results are proposed in the context of the simple BP model, with 5 parameters. The Form Factor modification of Eq.~(\ref{eq:bp7}) may introduce some changes, but, if we assume the parameter $t_0$ to asymptote to a constant value ( of $t_0\simeq 0.71\ GeV^2$),  its introduction will not spoil the simple relations of Eqs.~(\ref{eq:sr0}) and (\ref{eq:sr1}).  { We point out that the ansatz $B(s)\sim (\ln s)^2 $ is asymptotically consistent with data, as discussed in \cite{Schegelsky:2011aa} and seen in Fig. \ref{g:lhcbp6}. However,   at non-asymptotic energies the parameter $B(s)$ may have a slower growth.
  
To estimate the energy dependence of the parameters of the non-leading term, i.e. $D(s)$ and $C(s)$, is more complicate. 
 An important consequence of Eq. ~(\ref{eq:rhovsc}) is that $\sqrt{C(s)}$, if at all, must increase less than $\ln s$, if both $\rho(s)$ and ${\hat \rho(s)}\sim (\ln s)^{-1}$ asymptotically. Namely
\begin{itemize}
\item If the first [the $\sqrt{A}$] term in the elastic amplitude indeed represents a $C = +$ vacuum term,
then
\begin{equation}
\label{4.3}
 \hat{\rho}(s) \to\ \frac{\pi}{\ln (s/s_0)}.
\end{equation}
\item If the Froissart-Martin bound is indeed saturated, then we have the Khuri-Kinoshita theorem according to which also
\begin{equation}
\label{4.4}
 \rho(s) \to\ \frac{\pi}{\ln(s/s_0)};\ [{\rm with\ the\ same\ coefficient}\ \pi].
\end{equation}
\item If both Eq.(\ref{4.3}) and Eq.(\ref{4.4}) are simultaneously true, then we must have that
\begin{equation}
\label{4.5}
[\sqrt{(C/A)}] \ln(s/s_0) \to \  0.
\end{equation}
\item The above precludes $\sqrt{C}$ from growing  asymptotically as $\ln(s/s_0)$, if $\sqrt{A} \sim\ [\ln(s/s_o)]^2$.
\end{itemize}   
In the logarithmic approximations we are using here, the simplest assumption, { albeit not  the only one}, satisfying the sum rules, the Froissart bound and  the Khuri-Kinoshita theorem  \ \cite{Khuri:1965zz}, is then
\begin{equation}
D(s)\sim \ln s \  \ \ \ \ \ \ \ \ \ \ \ \ \ \sqrt{C(s)}\sim constant 
\end{equation}
In other terms, when the Khuri-Kinoshita asymptotic betaviour for the real part of the amplitude is satisfied,
% at present energies, 
one can choose the amplitude $C(s)\rightarrow constant$ and the sum rules dictate a normal Regge-like  behavior for the slope of non-leading term, $D(s)$.
 However there are some caveats and subtleties to be aware of:
\begin{itemize}
\item (i) the phenomenology presented for $pp$, and $\pbarp $ scattering as well, shows $\sqrt{C(s)}$ to increase very rapidly from ISR to LHC7, hence a constant behavior over this energy range is not observed (see Table \ref
{t:lhcbp2}).
\item (ii) For a large range of energy interval $\rho (s) \sim constant$ [average value $0.12$] and thus over the same range of interval $C(s)$ may increase in order to  keep
$SR_0 \sim 0$. This seems to be borne out by the phenomenology.
\end{itemize}
Thus it is quite possible that, at least in the energy range in which $\rho(s)\sim constant$,  $\sqrt{C} \sim \ln(s/s_0)$. Unfortunately, with present data, no unique limit can be prescribed. We shall thus resort to an empirical parameterization for $\sqrt{C(s)}$, as shown shortly below.

\subsection{Phenomenological results for the parameters}
In this section we propose an empirical description of the differential elastic $pp$ cross-section to be used at LCH8 and LHC14.  This parametrization follows Eq. (\ref{eq:bp7}) and refines the one proposed in \cite{Grau:2012wy}, presenting an optimal description of the very small $-t$ value, in addition to the already mentioned good description of the dip and the tail.

Following the discussion in the paper, and fits to ISR and LHC7 data, we propose the following asymptotic parametrization:
\begin{eqnarray}
4  \sqrt{\pi A(s)}(mb)=47.8-3.8 \ln s+0.398 (\ln s)^2\label{eq:Aparam}\\
B(s)(GeV^{-2})=11.04+0.028 (\ln s)^2-\frac{8}{0.71}=-0.23+0.028 (\ln s)^2 \label{eq:Bparam}\\
4 \sqrt{\pi C(s)}(mb)=\frac{9.6 -1.8 \ln s +0.01( \ln s)^3}{1.2+0.001(\ln s)^3}\label{eq:Cparam}\\
D(s)(GeV^{-2})=-0.41+0.29 \ln s \label{eq:Dparam}
\end{eqnarray}
The parametrization for $C(s)$ is empirical, $\sqrt{A(s)}$ and $B(s)$ follow asymptotic maximal energy saturation behavior, $D(s)$ shows normal Regge behavior.
In Fig. \ref{fig:param} we  plots   the parametrizations from   Eqs. (\ref{eq:Aparam}), (\ref{eq:Bparam}), (\ref{eq:Cparam}), (\ref{eq:Dparam}), and indicate  the results of the fit to the elastic differential cross-sections for $pp$  data (black dots). The red dots indicate the value of the parameters obtained when { fitting $\pbarp$
data with the $mBP2$ model. We leave a discussion of this model for the $\pbarp$ case to a forthcoming paper.}
  Notice that ${\bar p}p$ data  were not used to determine  the parametrization given in Eqs.(\ref{eq:Aparam}),(\ref{eq:Bparam}),(\ref{eq:Cparam}),(\ref{eq:Dparam}). 
\begin{figure}[H]
\includegraphics[width=4cm,height=4cm]{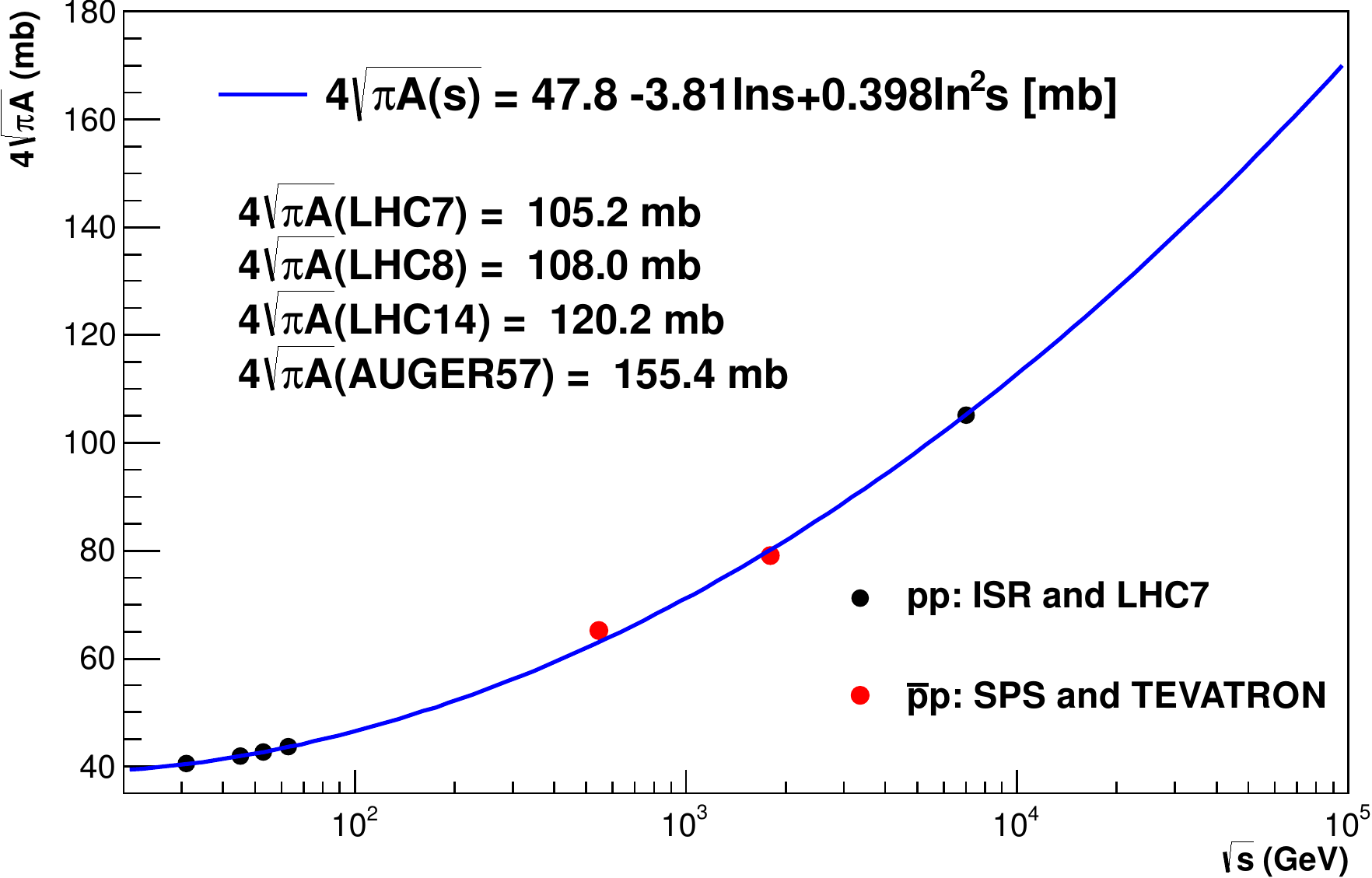}
\includegraphics[width=4cm,height=4cm]{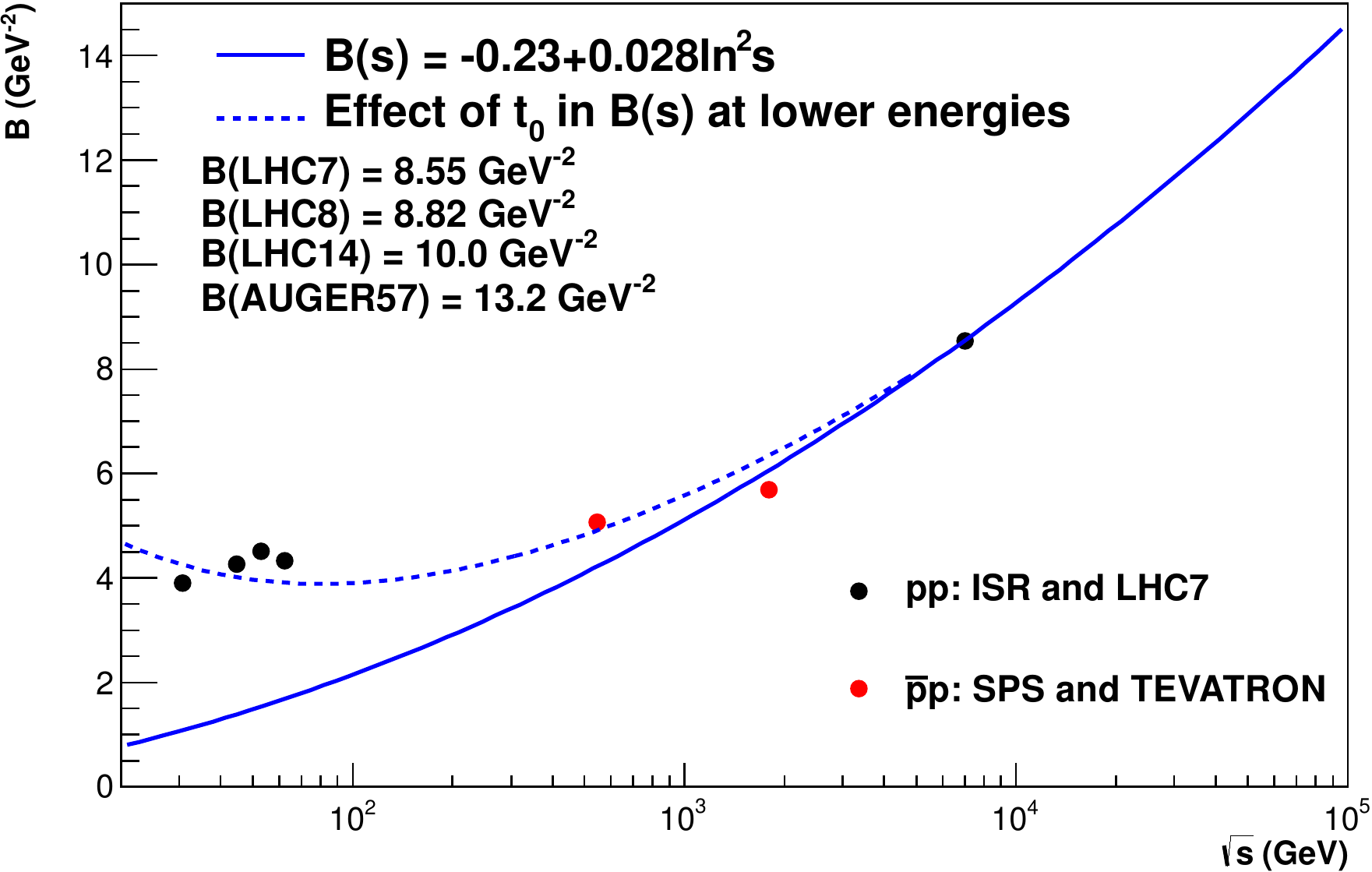}
%FIT_Beff_MAY17-D}
\includegraphics[width=4cm,height=4cm]{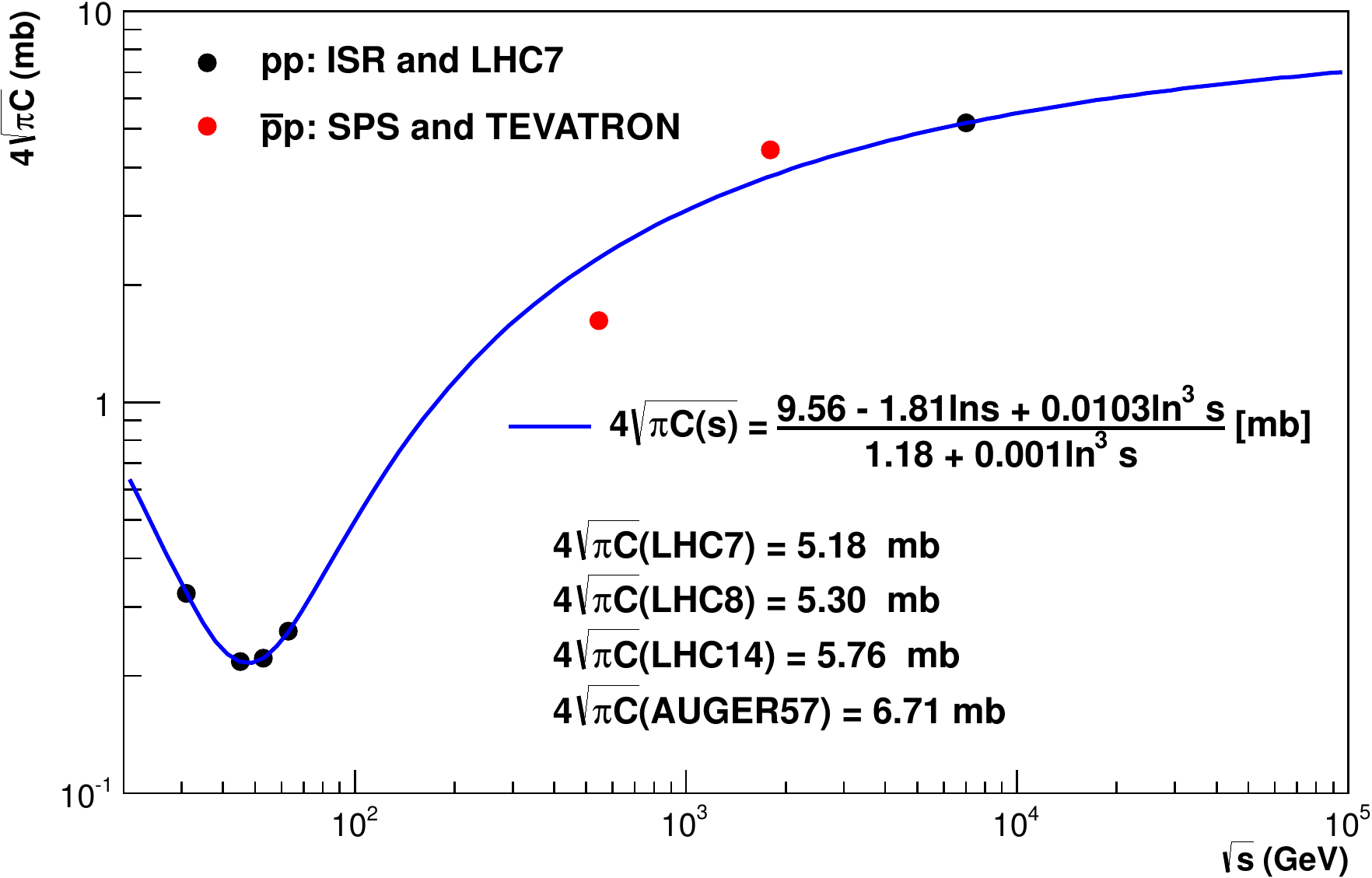}
\includegraphics[width=4cm,height=4cm]{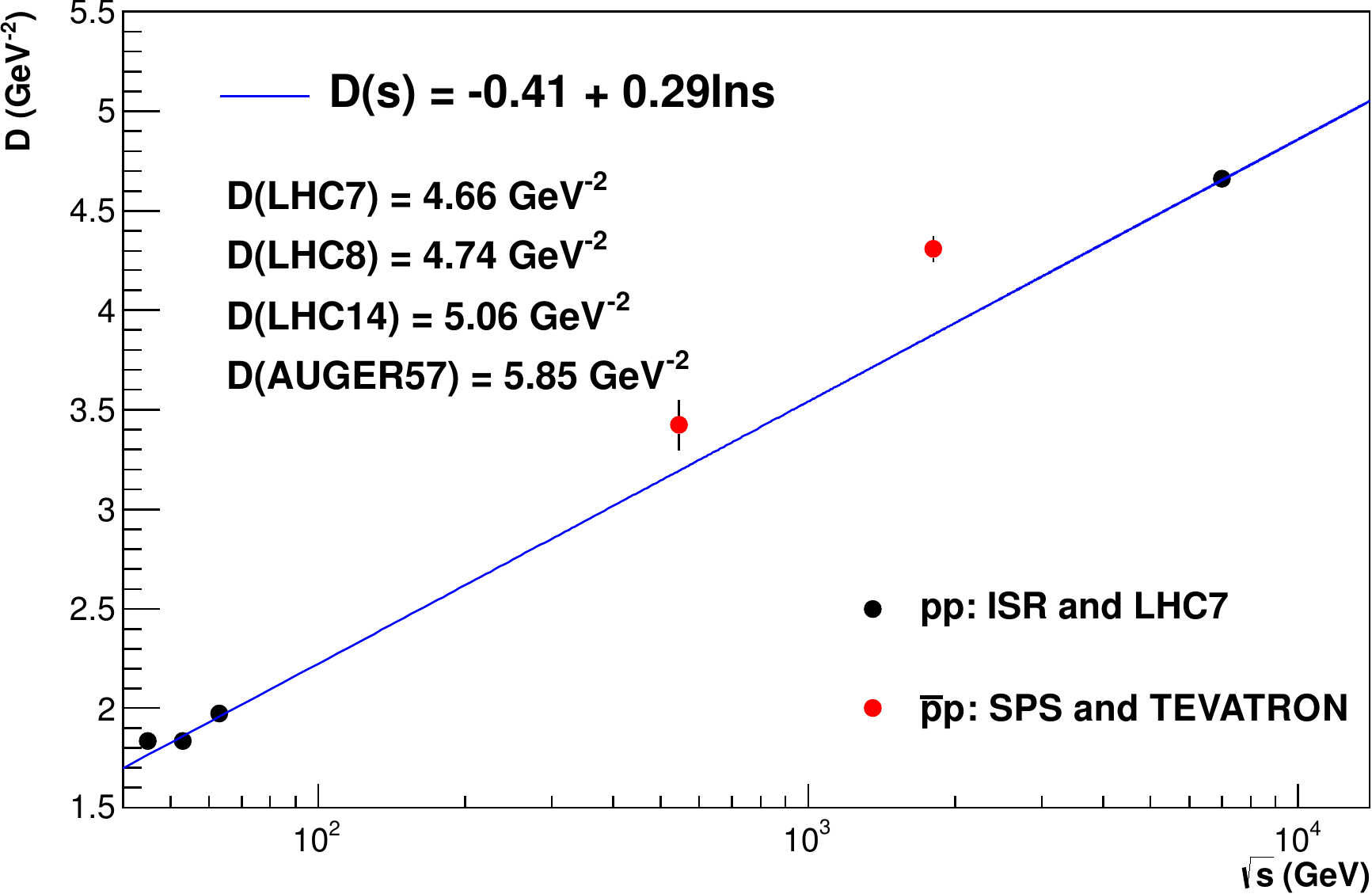}
\caption{Energy behavior of the model parameters $\sqrt{A(s)},\ B(s),\ \sqrt{C(s)}, D(s)$ described in the text. Black dots correspond to parameter values determined by fit to $pp$ scattering data. Red dots correspond to the parameter values fitting  ${\bar p} p$ data, and were not used to determine  the parametrization given in Eqs.(\ref{eq:Aparam}), (\ref{eq:Bparam}), (\ref{eq:Cparam}), (\ref{eq:Dparam}). }
\label{fig:param}
\end{figure}

\subsection{The position of the dip}
Although the phase $\phi$ is consistent with a constant as the energy increases, its value fluctuates. In the range $\sqrt{s}=53-7000 \ GeV$, the fits for $pp$ and $\pbarp$ indicate $\phi \simeq 2.7-2.8\ rad$. We note that the value used for $\phi$ influences the position and depth of the dip. In order to choose a value for $\phi$, we then study how the dip moves as a function of energy. The simplest asymptotic assumption for the dip position as a function of energy is to assume geometrical scaling, namely
$t \sigma_{total}\sim constant$.
In the maximal saturation model, in which $\sigma_{total}\sim (\ln s)^2$,  one can then parametrize the dip position as
\be
t_{dip}=-\frac{a}{1+b (\ln s)^2}\label{eq:tdipgsyogi}
\ee
In Fig. \ref{fig:tdip} we compare data for the position of the dip in $pp$ scattering with a parametrization obtained from   Eq. (\ref{eq:tdipgsyogi}) and with other predictions from amplitudes obeying  geometrical scaling as discussed in  \cite{Bautista:2012mq}}. A linear logarithmic fit is also shown for comparison.
\begin{figure}[H]
\centering
\includegraphics[width=10cm,height=8cm]{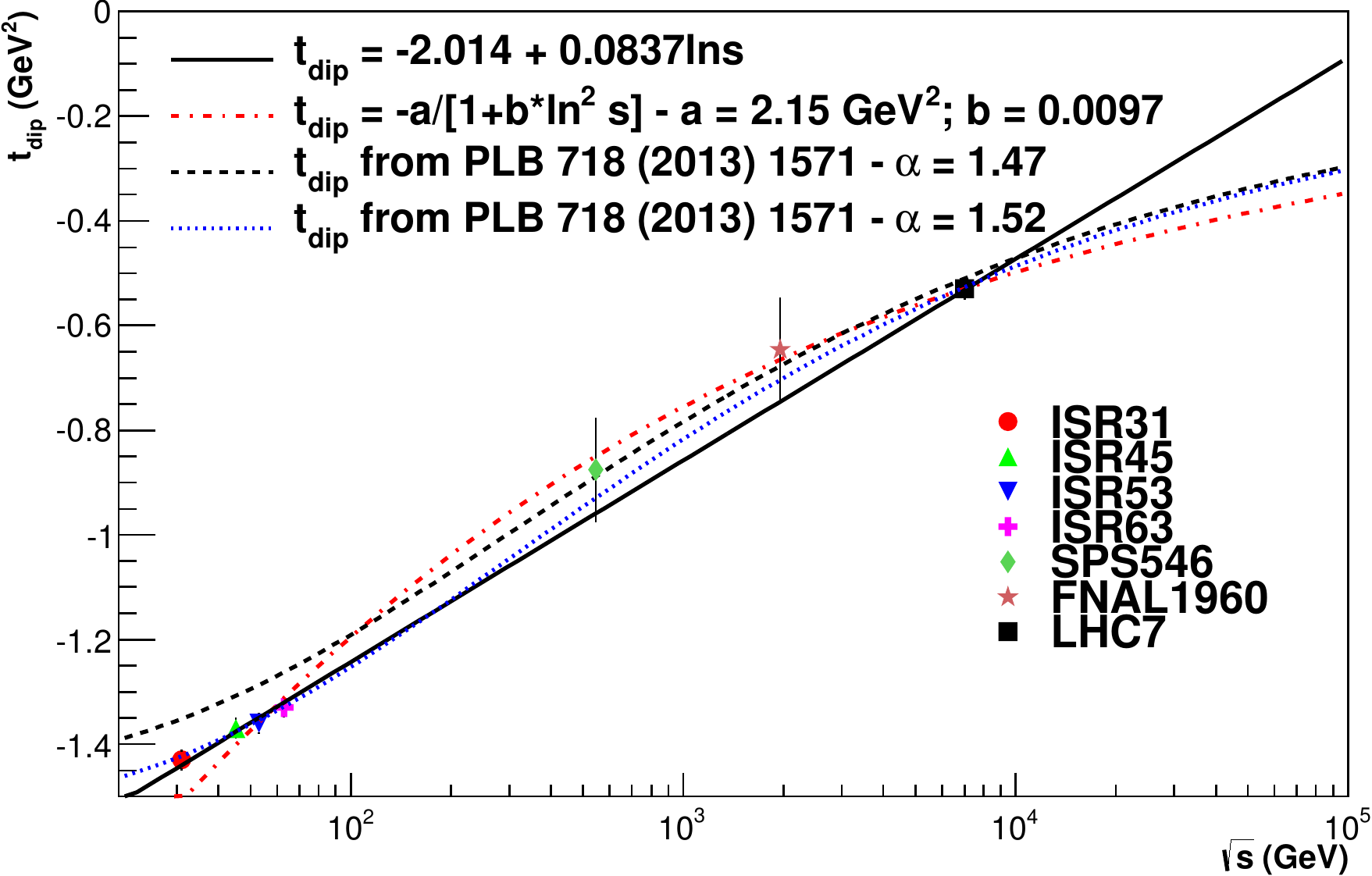}
\caption{Experimental values for the position of the dip in $pp$ and $\pbarp$ elastic scattering vs.  models suggested by geometrical scaling \cite{Bautista:2012mq} or a simple logarithmic energy rise.}
\label{fig:tdip}
\end{figure}   
Using these different possibilities, one can calculate the position of the dip at LHC8 and LHC14, as shown in Table \ref{tab:tdip}. In this table, $GS1$ refer to the parametrization of Eq.~(\ref{eq:tdipgsyogi}), $GS2$ and $GS3$ to different applications of the geometrical scaling model of Ref. \cite{Bautista:2012mq}. The geometrical scaling values are in good agreement with recent predictions  for the dip position at LHC14  from \cite{Bourrely:2012hp,Ferreira:2012zi}. 
\begin{table}[H]
\caption{Dip position from $\sqrt{s} = 8$ TeV onwards as predicted by geometrical scaling models and simple linear logarithmic evolution.}\label{tab:tdip}
\vspace*{0.2cm}
\centering
\begin{tabular}{ccccc}
\hline\hline $\sqrt{s}$ (TeV) & $|t|_{dip}^{LIN}$ & $|t|_{dip}^{GS1}$ & $|t|_{dip}^{GS2}$ & $|t|_{dip}^{GS3}$ \\ 
\hline 8 & 0.510 & 0.518 & 0.495 & 0.511 \\ 
\hline 14 & 0.417 & 0.471 & 0.439 & 0.452 \\ 
\hline\hline 
\end{tabular} 
\end{table}

\section{Predictions for LHC8 and LHC14 and the Black Disk limit}
\begin{figure}[H]
\begin{center}
\includegraphics*[width=8cm,height=6.5cm]{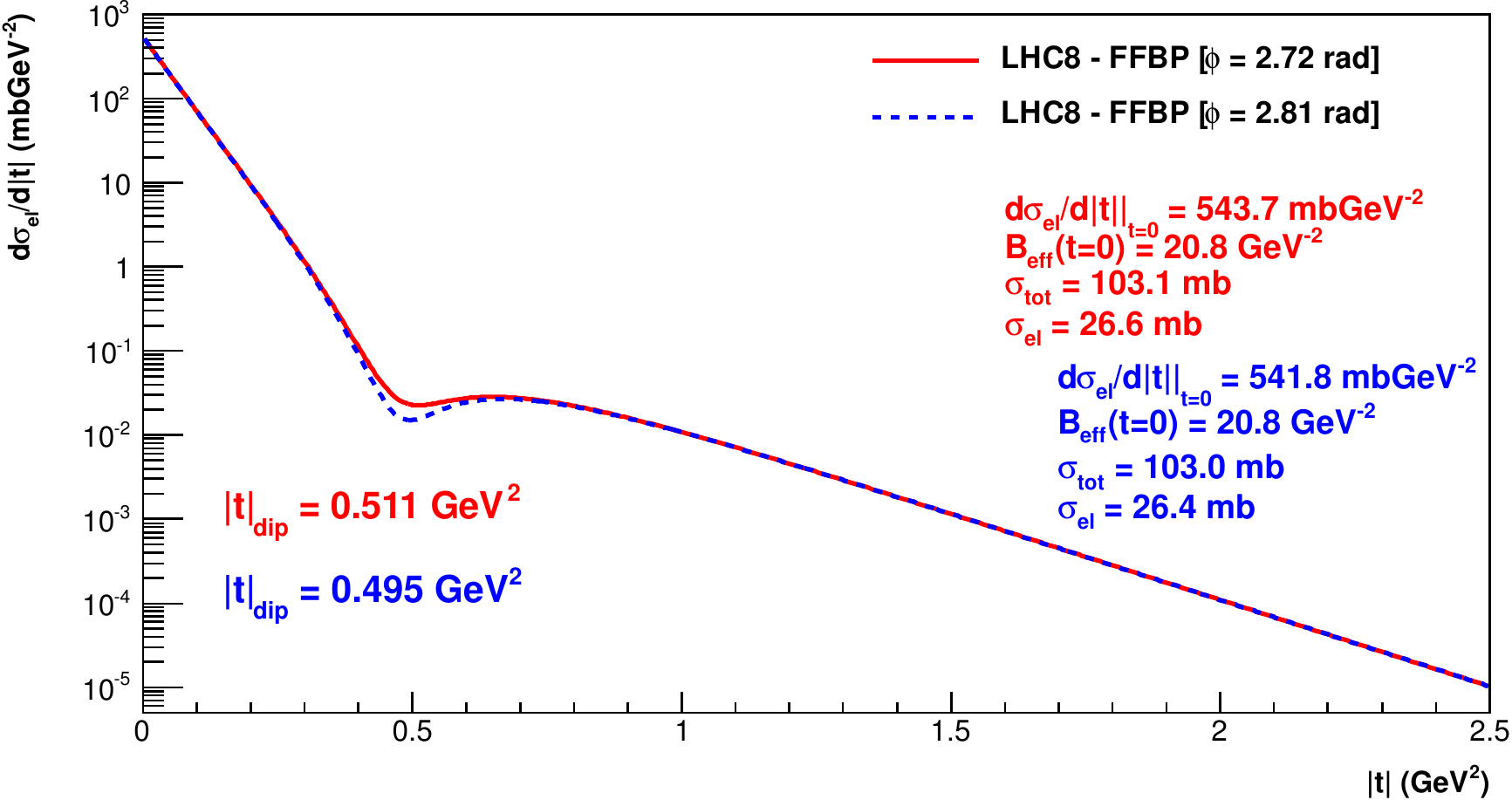}
\includegraphics*[width=8cm,height=6.5cm]{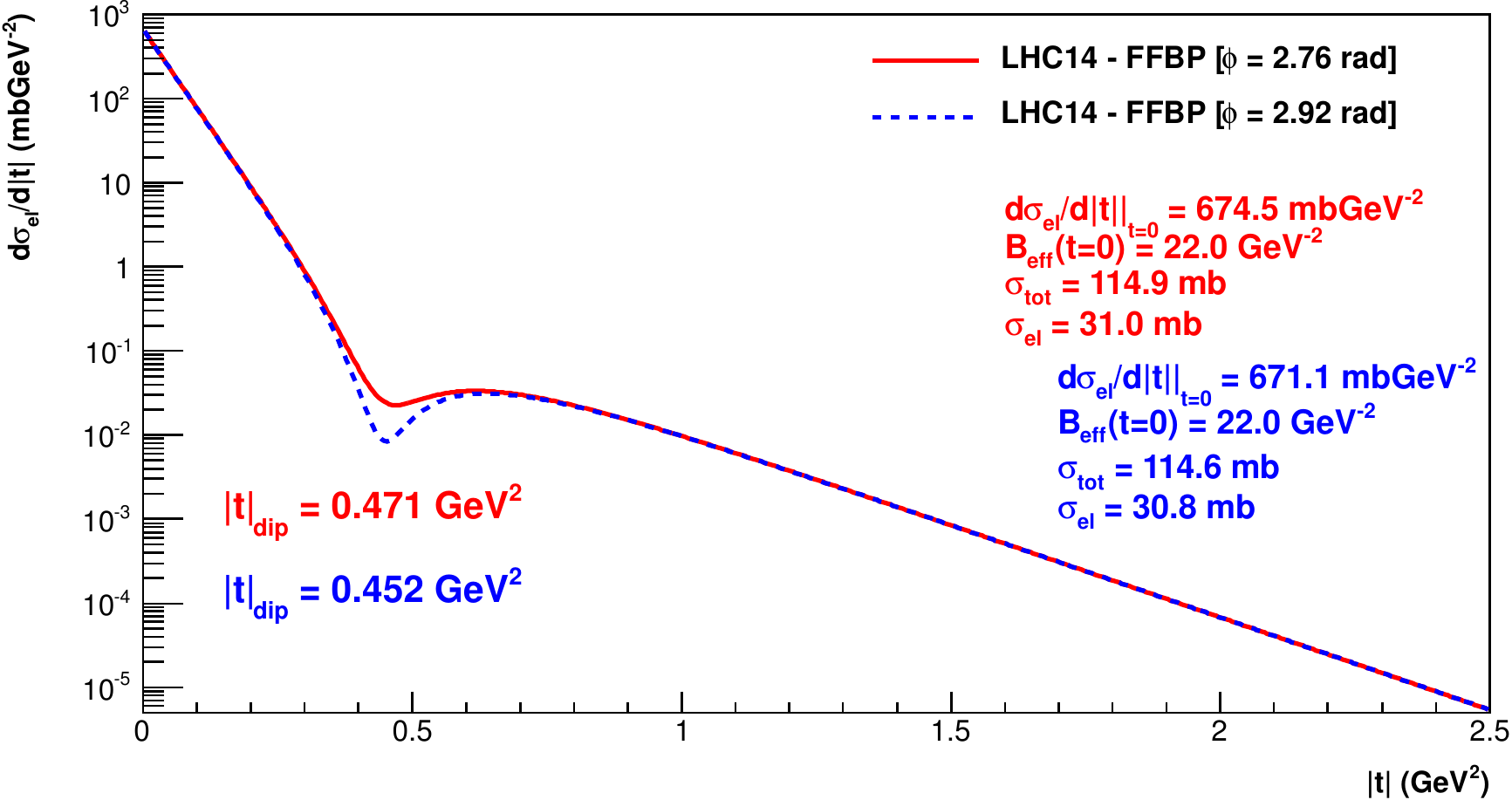}
\caption{  $mBP2$ model predictions for  the differential elastic cross section at LHC8 and LHC14 in a maximal energy saturation model, $\sigma_{total} \sim (\ln s)^2$.}
\label{g:lhcbp11}
\end{center}
\end{figure}

We are now in a position to predict the $t$-dependence of the elastic differential cross-section in $pp$ scattering at higher LHC energies, using the empirical asymptotic model described in the previous section.  
In Fig.\ref{g:lhcbp11} we  show these predictions for $pp$ elastic differential cross-sections  at LHC8 and LHC14. 
This model does not include  a second dip, or a wiggle, as in many eikonal models, such as for instance seen in \cite{Block:2012ym}. On the other hand, at present, at LHC7, in the interval $0<-t<2.5\ GeV^2$ TOTEM data do not allow to establish  the presence of a second dip or wiggle.
Finally, the dotted and full line correspond to different values of the phase $\phi$ and the figures confirm the sensitivity of the dip depth and position to  the chosen value for the phase $ \phi$.
{We now turn to higher energies and consider one favorite test of asymptotia, namely the black disk limit. As also noticed in \cite{Block:2012nj}, present data from LHC7 indicate that we are still far from this limit. The question is : how far? 

Using  the  energy parametrization discussed in the previous section, an  approximately constant scale  $t_0$ and  a band of values for $\phi$, we obtain  the result shown in Fig.\ref{fig:rel}. We notice that this ratio is in agreement with AUGER results \cite{Collaboration:2012wt}. Moreover, the asymptotic behavior is dictated by the Sum Rules, which reinforcing the condition of total absorption of partial waves, lead to the saturation of the black disk limit, i.e. $R_{el}\rightarrow 1/2$ as $s\rightarrow \infty$. From the parameters presented in Table \ref{tab:param}, we estimate that $R_{el}\simeq 1/2$ at $\sqrt{s} \simeq 10^{10}$ GeV (corresponding to the energy in the lab frame $E\simeq 10^{20}$ GeV), i.e. at energies typically  larger than the Planck scale.  
\vspace*{0.2cm}
\begin{figure}[H]
\begin{center}
\includegraphics[width=12cm,height=6.5cm]{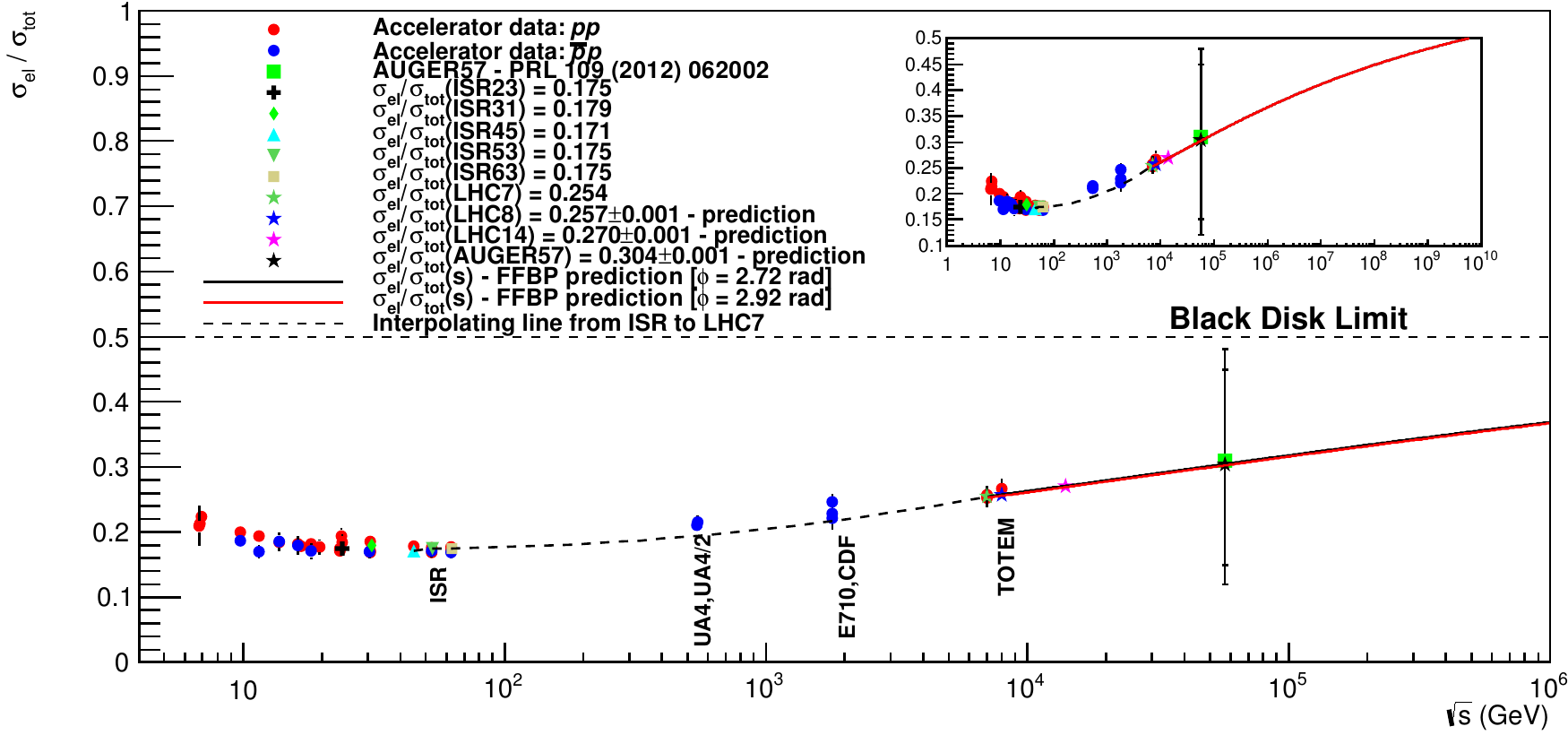}
\caption{Experimental data from accelerators for the ratio $R_{el}=\sigma_{elastic}/\sigma_{total}$ as compiled from \cite{Antchev:2013gaa,Antchev:2012prep8TeV,Fagundes:2012rr} and this model expectations. The AUGER datum has been extracted from the ratio $\sigma_{inel}/\sigma_{total}$ at $\sqrt{s} = 57$ TeV, as coming from   estimates presented in \cite{Collaboration:2012wt}. For this point, the \textit{inner bars} comprise only statistical and systematic uncertainties combined quadratically and the \textit{outer bars} incorporate the total uncertainty, with errors from Glauber calculations also summed in quadrature.
Inner bars:   $R_{el}^{\textit{stat+sys}}(57 TeV) = 0.31^{+0.14}_{-0.16} $,
outer bars:   $R_{el}^{\textit{stat+sys+Glauber}}(57 TeV) =
0.31^{+0.17}_{-0.19} $.}
\label{fig:rel}
\end{center}
\end{figure}  
As expected, the ratio $\mathcal{R}_{el}$ is less sensitive to variations in $\phi$, since the  contribution arising from the dip region to the integrated elastic cross section is minimal. Therefore, notwithstanding  the observable effect in the elastic differential cross section, shown in  Fig. \ref{g:lhcbp11}, the predictions of this model for different values of $\phi$ lead to practically overlapping curves.}

\begin{table}[H]
\centering
\caption{  Values of $mBP2$ parameters used in the predictions at LHC8, LHC14 and AUGER57 and the ratio $\mathcal{R}_{el}$ at each c.m. energy. In all cases the values of $t_{0}$ has been frozen at 0.71 $GeV^{2}$ and bands for $\phi$ were considered. These bands  determine the uncertainty in the predictions for the ratio.}
\vspace*{0.3cm}
\begin{tabular}{c|c|c|c|c|c||c}
\hline \hline $\sqrt{s}$ (TeV) & $A$ (mbGeV$^{-2}$) & $B$ (GeV$^{-2}$) & $C$ (mbGeV$^{-2}$) & $D$ (GeV$^{-2}$) & $\phi$ (rad)& $\sigma_{el}/\sigma_{tot}$ \\ 
\hline 8.0 & 596 & 8.8  & 1.44 &  4.7 & 2.72$-$2.81 & 0.257 $\pm$ 0.001 \\ 
\hline 14 & 739 & 10.0 & 1.70 & 5.1  & 2.76$-$2.92 & 0.270 $\pm$ 0.001 \\ 
\hline 57 & 1233 & 13.2  & 2.30 & 5.9 & 2.72$-$2.92 & 0.304 $\pm$ 0.001\\ 
\hline \hline
\end{tabular}
\label{tab:param}
\end{table}

\section*{Conclusions}
We have shown that  the $pp$  differential elastic cross-section   in the range measured by the TOTEM experiment at LHC can be parametrized  through  two exponentials and a phase,  provided the first term is modified by  a multiplicative  factor to optimize the description of the forward peak.  Two different modifications are proposed. For the model with a proton  form factor to modify the $-t\simeq 0$ behavior, we extract predictions at LHC8 and LHC14, and calculate  the ratio of the elastic to the total cross-section up to and beyond $\sqrt{s}=57\ TeV$.

{ The parametrization of $pp$ elastic cross-section data presented in this paper is not meant to be exact, rather to indicate how to break up the amplitude in a set of building blocks, and apply this dissection  to the data as the energy increases.  This parameterization addresses the following basic elements:
\begin{itemize}
\item the value of the differential cross-section at $t=0$, namely the optical point value
\item a rapid  decrease, characterized by a slope, which, between $-t=0$ and the dip, is not a constant
\item the occurrence of a dip in $pp$ at all energies from ISR to LHC
\item an exponential decrease after the dip, with a non-leading slope and an amplitude much smaller than before the dip.
\end{itemize}
This behavior is described by an empirical model, with two amplitudes, two different slopes, a phase and  the 
proton form factor to multiply the amplitudes.} This  empirical model  might help us to understand the elastic $pp$ differential cross-section \cite{Soffer:2013tha}. It describes the data well and, as such, can be used by model builders and experimentalists alike. 

The interpretation of the model is in parts straightforward, but not completely.  In our previous analysis of TOTEM data for the elastic differential cross-section \cite{Grau:2012wy}, we have commented on the physical meaning of the model. Our considerations were that the two terms in the amplitude receive  contributions from different charge conjugation processes, the first term purely from $C=+1$, the second 
non-leading term has contributions { from both $C=\pm 1$ terms,  which, { at high energy}, render $\phi\neq \pi,\ \pi/2$. }The energy behavior of the leading amplitude $A(s)$ is consistent with many eikonal models, but the  exponential behavior in the momentum transfer before and after the dip is not, and it is probably due to rescattering effects in the final state. On the other hand, the modification of the model with a form factor which reproduces the proton electromagnetic form factor at high energy, suggests the need to include rescattering effects within  each colliding hadron, namely the probability that the proton does not break up as the momentum transfer increases.

\section*{Acknowledgments}
We thank L. Jenkovszky, M. J. Menon and J. Soffer for useful  discussions. AG acknowledges partial support by Spanish MEC (FPA2010-16696,
AIC-D-2011-0818) and by Junta de Andalucia (FQM 03048, FQM 6552, FQM 101).
DAF acknowledges the S\~ao Paulo Research Foundation (FAPESP) for financial
support (contract: 2012/12908-4).
\begin{appendices}
\appendix
\section{Two-pion threshold effects on the BP model: {\it mBP1}\label{app:2pion}}
\begin{figure}[H]
\begin{center}
\includegraphics*[width=8.5cm,height=7.5cm]{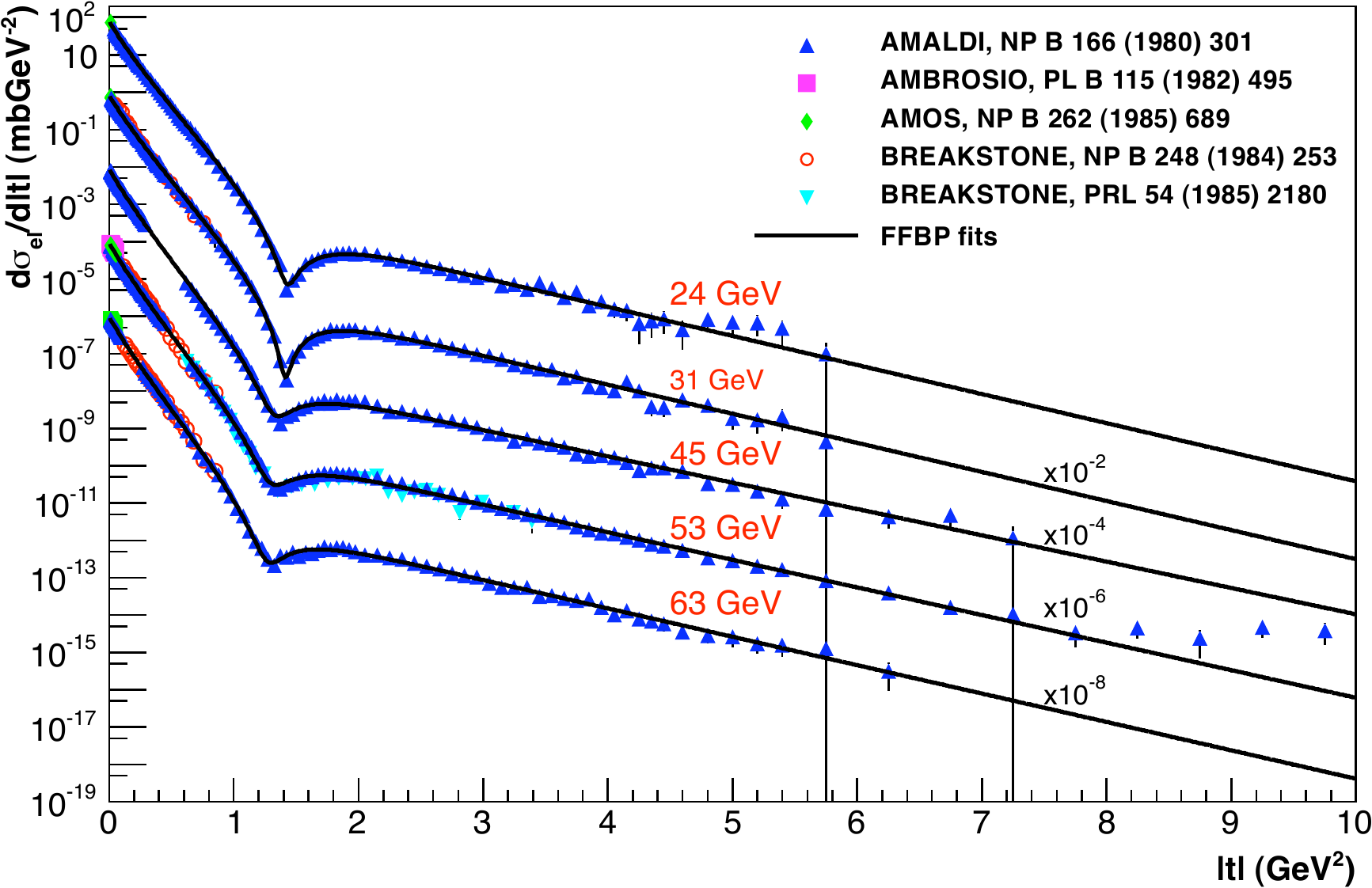}
\includegraphics*[width=8.5cm,height=7.5cm]{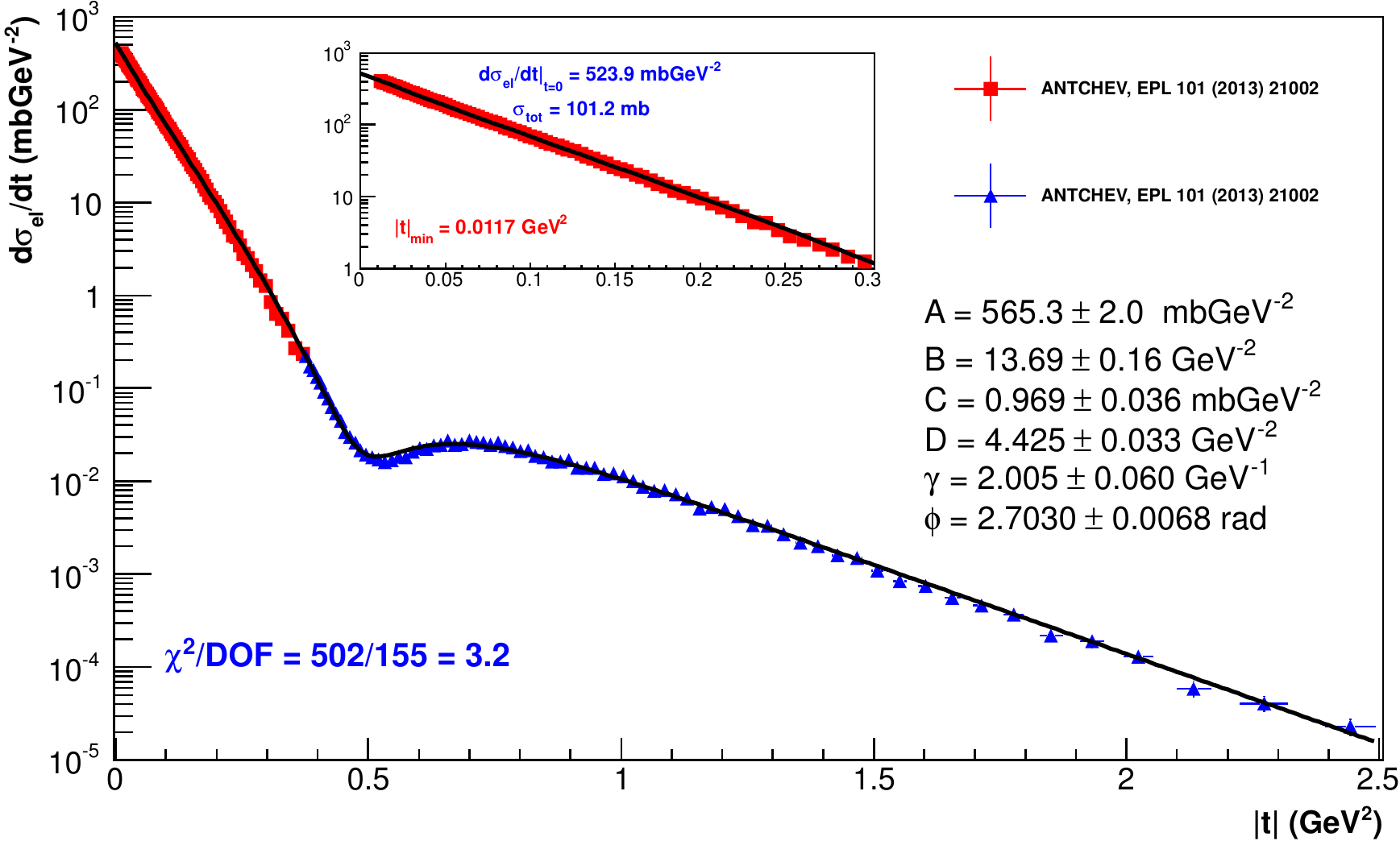}
\caption{Fits to the ISR and LHC7 data sets with model $mBP1$.}
\label{g:lhcbp3}
\end{center}
\end{figure}

\begin{figure}[H]
\begin{center}
\includegraphics*[width=8cm,height=6cm]{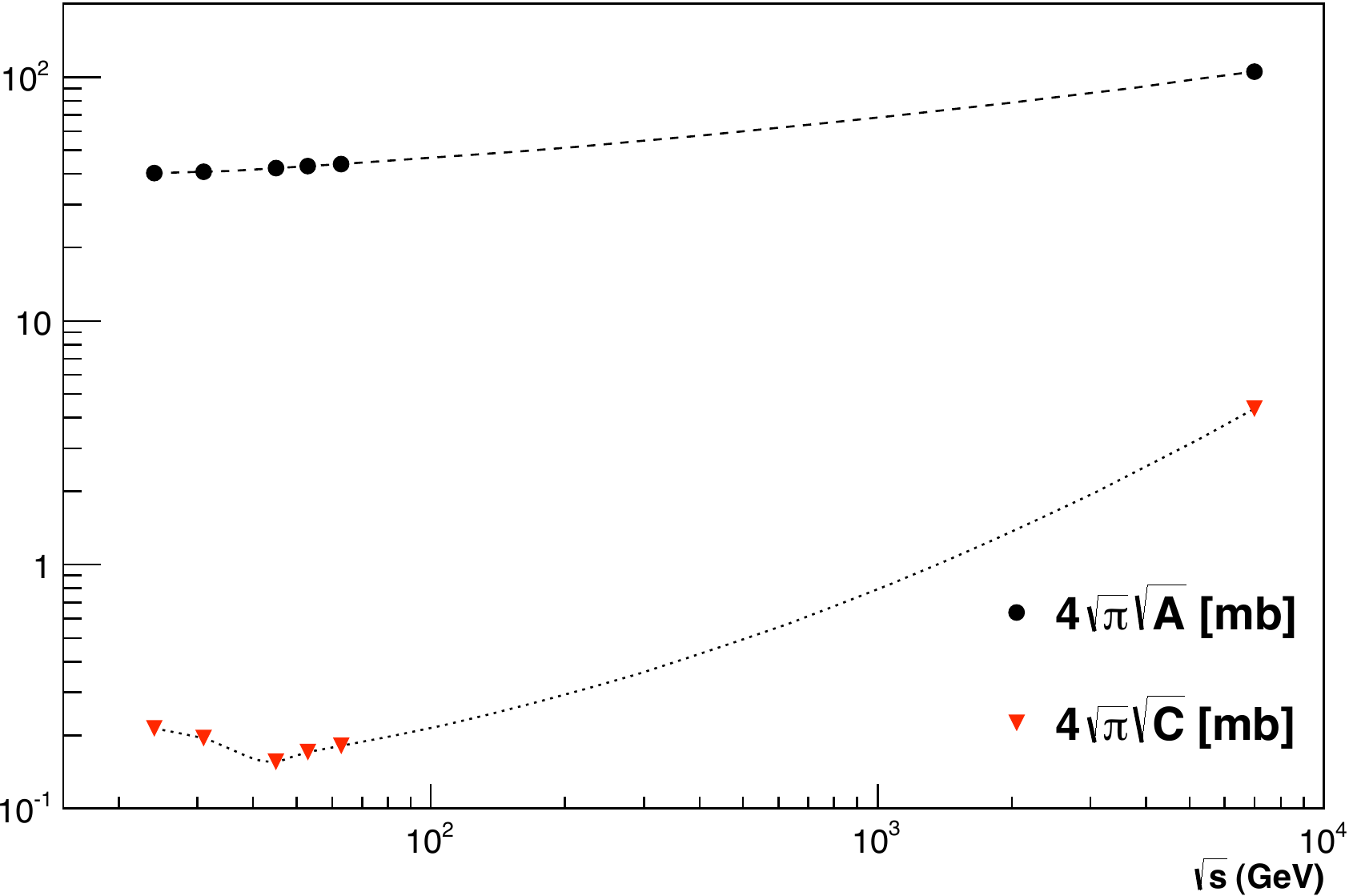}
\includegraphics*[width=8cm,height=6cm]{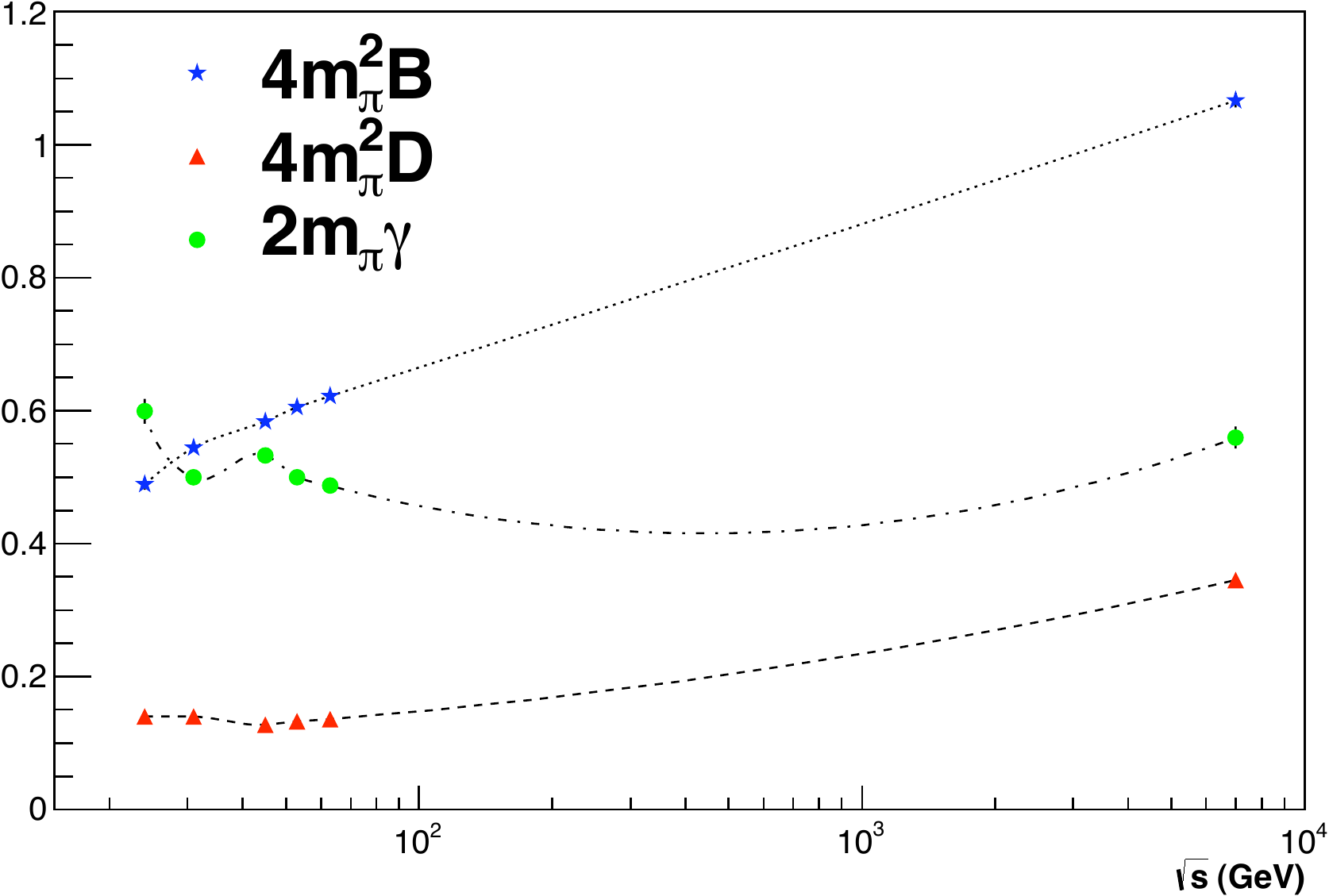}
\caption{Energy behavior of parameters from the $mBP1$ model.}
\label{g:lhcbp_pars_mbp1}
\end{center}
\end{figure}

We discuss here a model where the very small $-t$ behavior 
 is influenced by the nearest  $t-$channel singularity of the scattering amplitude. In this model, which we call $mBP1$,
\begin{equation}
{\cal A}(s,t)=i[\sqrt{A(s)}e^{B(s)t/2}G(s,t)+ e^{i\phi} \sqrt{C(s)}e^{D(s)t/2}],\label{eq:bp1}
\end{equation}
with $G(s,0)=1$
in order not to spoil the good description of the dip by Eq.~(\ref{eq:bp0}), as discussed in the text.
Such factor  would arise from    the contribution of  the \textit{two-pion loop}  in the Pomeron trajectory  as originally proposed in \cite{Anselm:1972ir} and  \cite{CohenTannoudji:1972gd},   and more recently discussed by Khoze, Martin and Ryskin \cite{Khoze:2000wk} and Jenkovszky  \cite{Jenkovszky:2011hu,Fiore:2008tp}.  In particular $\alpha_P(t)$, at very small $t$,  should include   a square root singularity at $t=4\mu^2$, with $\mu$ the pion mass.
Mindful of such possibilities, we have applied the following correction to the first  term  of Eq. (\ref{eq:bp0}), namely we shall use 
\begin{eqnarray}
G(s,t) = e^{-\gamma(s)(\sqrt{4\mu^{2}-t}-2\mu)},\label{eq:bp2}
\end{eqnarray} 
 with $\gamma (s)$ a free parameter. Being applied to the near-forward region, such term shall influence the small $|t|$ behavior of elastic differential cross section,  producing a changed curvature in the effective slope $B_{eff}(s,t)$ in  this region. 
The original expressions for the total cross section and the optical point remain  unchanged, but  
the  modification of the model of Eq.~(\ref{eq:bp0}) given by Eqs.~(\ref{eq:bp1}, \ref{eq:bp2})  introduces an additional $t$-dependence in the first term, through a square root, and hence a sixth parameter. Using the modified BP model of Eqs. (\ref{eq:bp1}, \ref{eq:bp2}) (henceforth called $ mBP1 $), we update our fits \cite{Grau:2012wy}  to LHC7 data samples as well as to the ISR data sets in the full range for $pp$ data with $\sqrt{s}= (23\div63)$ GeV, as  displayed in  Table \ref{t:lhcbp1} and Fig. \ref{g:lhcbp3}. ISR data sets used in the fits  comprise the data collection by Amaldi and Schubert \cite{Amaldi:1979kd} and all experimental information available from 1980 onwards \cite{Amos:1985wx,Breakstone:1984te,Breakstone:1985pe,Ambrosio:1982zj}. This table shows that this modification  gives  an acceptable  statistical description  from the optical point to the full $|t|$ range. 
\begin{table}[H]
\caption{
Values of free fit parameter $A, B, C, D, \gamma$ and $\phi$ at each energy analyzed. $A$ and $C$ are expressed in units mbGeV$^{-2}$, $B$ and $D$ in units GeV$^{-2}$, $\gamma$ in units GeV$^{-1}$ and $\phi$   in radians.}
\vspace{0.2cm}
\centering
\renewcommand{\arraystretch}{1.5}
{\scriptsize
\begin{tabular}{|c|c|c|c|c|c|c||c|}
\hline \hline 
$\sqrt{s}$ (GeV) & $A$ & $B$ & $C$ ($\times$10$^{-3}$)  & $D$  & $\gamma$  & $\phi$ & $\frac{\chi^{2}}{\rm DOF}$\\ 
\hline 24 &  82.8 $\!\pm\!$ 1.0 & 6.3 $\!\pm\!$ 0.1&  2.3 $\!\pm\!$  0.2  &  1.79 $\!\pm\!$ 0.04
  &  2.15 $\!\pm\!$ 0.07  &  2.94 $\!\pm\!$ 0.01 &  $\frac{200}{134-6}=1.1$ \\  
\hline 31 & 85.1 $\!\pm\!$ 0.2   &  6.99 $\!\pm\!$ 0.06  &  1.9 $\!\pm\!$ 0.1  & 1.79 $\!\pm\!$  0.02  
& 1.79 $\!\pm\!$ 0.03   & 3.02 $\!\pm\!$ 0.01 &  $\frac{310}{206-6}=1.6$ \\ 
\hline 45 & 91.5 $\!\pm\!$ 0.2   &  7.51 $\!\pm\!$ 0.05  & 1.18 $\!\pm\!$ 0.06   &  1.62 $\!\pm\!$ 0.02  
& 1.92 $\!\pm\!$ 0.03   & 2.73 $\!\pm\!$ 0.02   & $\frac{801}{207-6}=4.0$ \\  
\hline 53 & 94.6 $\!\pm\!$ 0.1   & 7.78 $\!\pm\!$ 0.05   & 1.49 $\!\pm\!$ 0.05   & 1.70 $\!\pm\!$ 0.01   &  1.79 $\!\pm\!$ 0.02  &  2.68 $\!\pm\!$ 0.01  & $\frac{1490}{319-6}=4.8$ \\ 
\hline 63 & 98.5 $\!\pm\!$ 0.2   & 7.98 $\!\pm\!$ 0.09   & 1.7 $\!\pm\!$ 0.1   & 1.75 $\!\pm\!$ 0.03  & 1.74 $\!\pm\!$ 0.04    & 2.75 $\!\pm\!$ 0.03   &  $\frac{332}{165-6}=2.1$ \\
\hline 7000 & 565 $\!\pm\!$ 2  & 13.7 $\!\pm\!$  0.2 &  970 $\!\pm\!$  40   & 4.43 $\!\pm\!$ 0.03 & 2.01 $\!\pm\!$ 0.06     & 2.703 $\!\pm\!$ 0.007  & $\frac{497}{161-6}=3.2$\\
\hline \hline
\end{tabular}}
\label{t:lhcbp1}
\end{table}
%\end{document} %%%%%%%%%%%%%%%%%%%%%%%%%%%%%qui%%%%%%%%%%%
In Fig. \ref{g:lhcbp_pars_mbp1} we present the energy dependence of fit parameters for $mBP1$ model . The continuous (dotted) lines in these figures are computer parametrizations drawn to guide the eye.

 We now  make two comments. Firstly, the square root factor is only used for the first term of the BP amplitude, as  this factor comes from the contribution of the pion loop to the 
leading vacuum term and 
it may not be present for the second non-leading term, which, { 
 for a generic $\phi$, 
 has contributions  also from $C=-1$ processes.} The second comment derives from an inspection of Table \ref{t:lhcbp1} and the energy dependence of the parameter $\gamma(s)$. This energy dependence, displayed in Fig. \ref{g:lhcbp_pars_mbp1} from ISR to LHC7, shows  a very slow increase, even compatible with a constant in energy, shedding  doubt on a straightforward interpretation of   this factor in terms of a small $t$ contribution to the Pomeron trajectory. 
 
 %\appendix

 The elastic cross section for this model from  Eqs. (\ref{eq:bp1}-\ref{eq:bp2}) is obtained as  
\begin{eqnarray}\label{eq:bp3}
\sigma_{el}(s) = \int_{-\infty}^{0} dt |\mathcal{A}(s,t)|^{2} = Ae^{4m_{\pi}\gamma}\mathcal{I}_{1} +
\frac{C}{D} + 2\sqrt{AC}e^{2m_{\pi}\gamma}\cos \phi \mathcal{I}_{3},
\end{eqnarray} 
where the integrals $\mathcal{I}_{1}$ and $\mathcal{I}_{3}$ are given as:
\begin{eqnarray}\label{eq:bp04}
\mathcal{I}_{1} &=& \int_{-\infty}^{0} dt\ e^{Bt-2\gamma\sqrt{4m_{\pi}^{2}-t}}, \\
\mathcal{I}_{3} &=& \int_{-\infty}^{0} dt\ e^{(B+D)t/2-\gamma\sqrt{4m_{\pi}^{2}-t}}.
\end{eqnarray}
An analytical evaluation can be obtained, used  the result:
\begin{eqnarray}\label{eq:bp5}
\mathcal{I}(\alpha,\beta,\delta) \equiv \int_{-\infty}^{0} dt e^{\alpha t-\beta\sqrt{\delta^{2}-t}}
= \frac{1}{\alpha}e^{-\delta\beta} - \frac{\beta \sqrt{\pi}}{2\alpha^{3/2}}Erfc[\sqrt{\alpha}
\left( \delta + \beta/2\alpha \right)]e^{\alpha\delta^{2}+\beta^{2}/4\alpha}, 
\end{eqnarray}
where $Erfc(x) = \frac{2}{\sqrt{\pi}} \int_{x}^{\infty} e^{-y^{2}}dy$ denotes the \textit{complementary error function}. Thus, from Eqs.~(\ref{eq:bp04} - \ref{eq:bp5}) it follows that:
\begin{eqnarray}\label{eq:bpelpion}
\sigma_{el}(s) &=& \frac{A}{B} + \frac{C}{D} + \frac{4\sqrt{AC}}{(B+D)}\cos\phi - \sqrt{\pi}\frac{A\gamma}{B^{3/2}}Erfc\left[\sqrt{B}\left(2m_{\pi}+\frac{\gamma}{B} \right)  \right]e^{4m_{\pi}^{2}(B+\gamma/m_{\pi})+\gamma^{2}/B} \nonumber\\
&-&\sqrt{8\pi}\frac{\sqrt{AC}\gamma\cos\phi}{(B+D)^{3/2}} Erfc\left[\sqrt{\frac{B+D}{2}}\left(2m_{\pi}+\frac{\gamma}{B+D} \right)  \right]e^{2m_{\pi}^{2}(B+D+\gamma/m_{\pi})+\gamma^{2}/2(B+D)}
\end{eqnarray}

In the  above expression one can see that  the contributions with positive sign  come from the simple BP amplitude, as one  can be easily checked by taking the limit $\gamma\rightarrow 0$. Thus, the presence of negative terms in Eq. (\ref{eq:bpelpion}), being due to $G(s,t)$, reflects the importance of modifying the first term of the original BP amplitude.

The sum rules for the elastic amplitude presented  in Ref. \cite{Grau:2012wy} can be applied to this model, and used to to check the saturation of the elastic amplitude at LHC energies.
One has:
\begin{eqnarray}
SR_{1} &=& \frac{1}{\sqrt{\pi}B}\sqrt{\frac{A}{1+\hat{\rho}^{2}}}+\frac{\sqrt{C}}{\sqrt{\pi}D}\cos\phi -\sqrt{\frac{\pi}{2}\frac{A}{1+\hat{\rho}^{2}}}\frac{\gamma}{B^{3/2}}Erfc\left[\sqrt{\frac{B}{2}}\left( 2m_{\pi} +\frac{\gamma}{B} \right)  \right]e^{2m_{\pi}^{2}(B+\gamma/m_{\pi})+\gamma^{2}/2B}; \label{eq:bp14}\\
SR_{0} &=& \frac{\hat{\rho}}{\sqrt{\pi}B}\sqrt{\frac{A}{1+\hat{\rho}^{2}}} -\frac{\sqrt{C}}{\sqrt{\pi}D}\sin\phi -\hat{\rho}\sqrt{\frac{\pi}{2}\frac{A}{1+\hat{\rho}^{2}}}\frac{\gamma}{B^{3/2}}Erfc\left[\sqrt{\frac{B}{2}}\left( 2m_{\pi} +\frac{\gamma}{B} \right)  \right]e^{2m_{\pi}^{2}(B+\gamma/m_{\pi})+\gamma^{2}/2B}.\label{eq:bp15}
\end{eqnarray} 
As above for the elastic cross section, the first two terms come from  the original BP amplitude and the input $G(s,t)$ produce the last term.
\vspace*{0.2cm}
\section{Other form factor modifications of the Barger and Phillips model\label{app:ffBP}}

We  examine here two more possible modifications of the Barger and Phillips model, complementary to the form factor modification of the first term,  presented in the text:
\begin{itemize}
\item the entire  BP amplitude is multiplied by a factor

%Eq06
\begin{equation}\label{eq:ff}
F^2_P=\frac{1}{(1-t/t_0)^4}
\end{equation}

with $t_0$ a free parameter,
namely

%Eq07
\begin{equation}
{\cal A}(s,t)=iF^2_P(t)[\sqrt{A(s)}e^{B(s)t/2}+e^{i\phi(s)}\sqrt{C(s)}e^{D(s)t/2}]\label{eq:bp6}
\end{equation}

\item both  terms of the BP amplitude are multiplied by a form factor (squared), but with difference scales,  $t_0$ and $t_O$, namely

%Eq09
\begin{equation}\label{eq:bp8}
{\cal A}(s,t)=i[F^2_P(t)\sqrt{A(s)}e^{B(s)t/2}+e^{i\phi(s)} F^2_O(t)\sqrt{C(s)}e^{D(s)t/2}]
\end{equation}
with 

%Eq10
\begin{equation}
F^2_O=\frac{1}{(1-t/t_O)^4}\label{eq:bp9}
\end{equation}
with $t_{0,O}$ free parameters.
\end{itemize}

\begin{figure}[H]
\begin{center}
\includegraphics*[width=6.5cm,height=5.0cm]{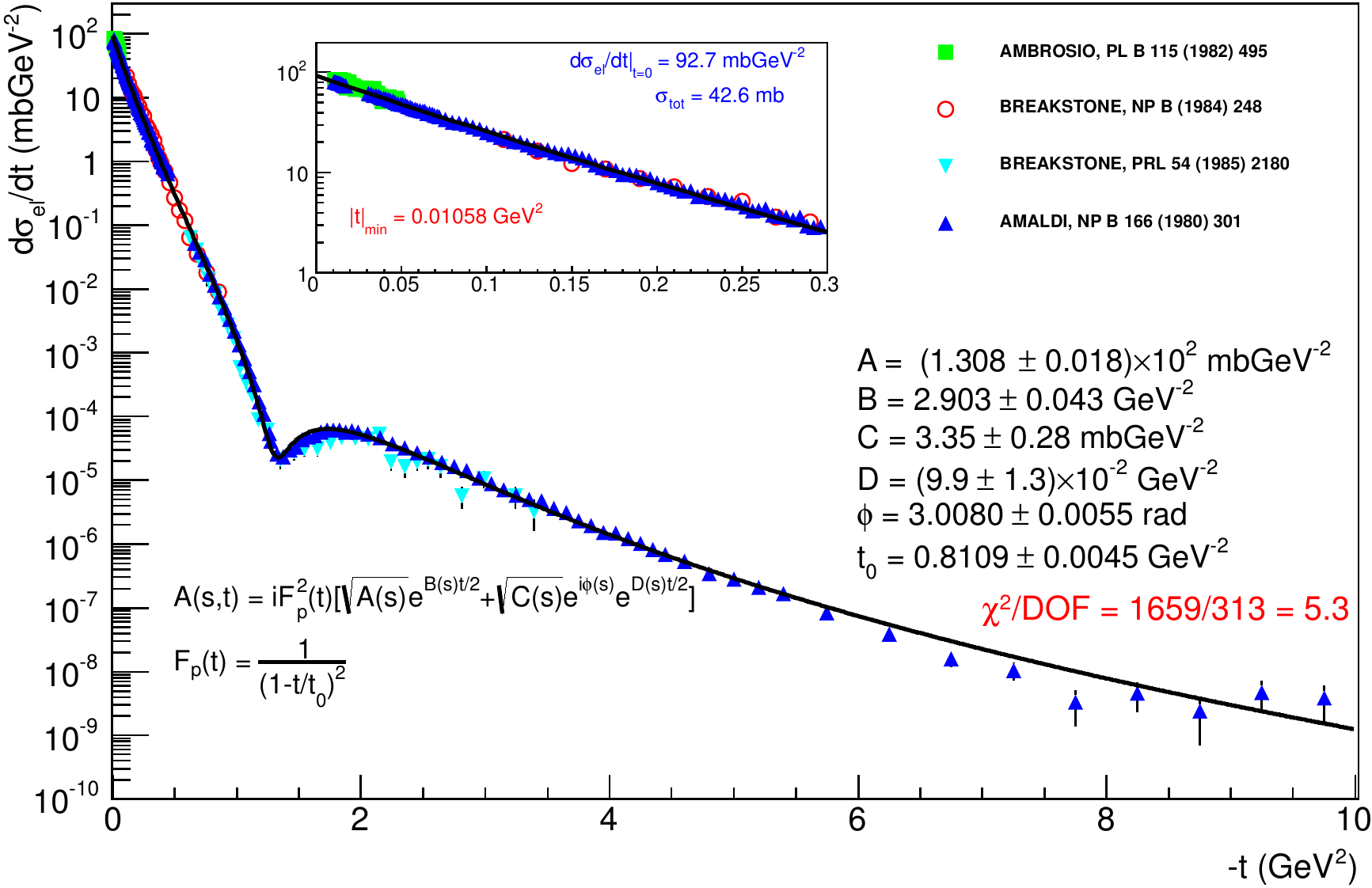}
\includegraphics*[width=6.5cm,height=5.0cm]{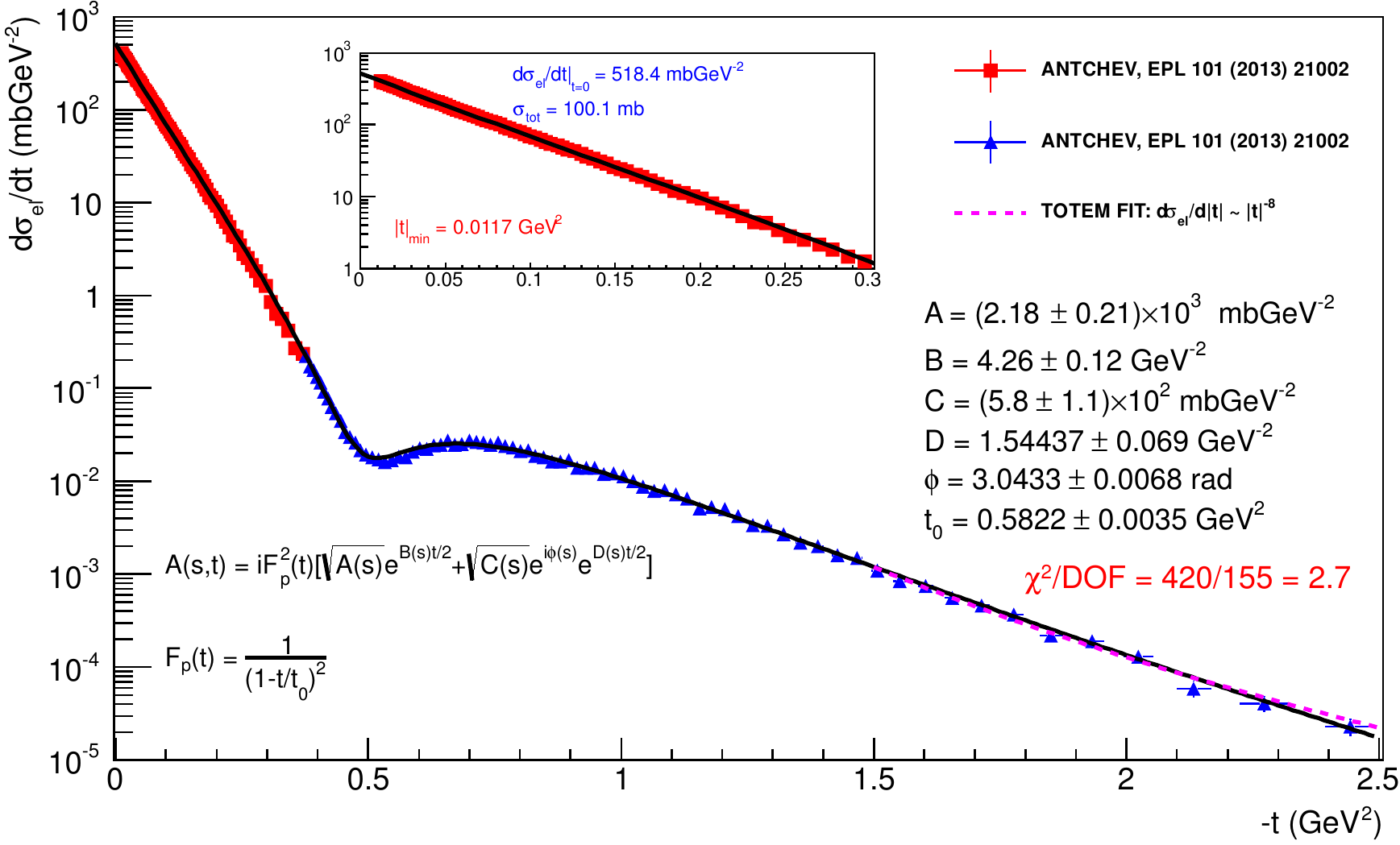}
%}
%LHC7_oBP_FF_GLOBAL}
%\includegraphics*[width=6.5cm,height=5.0cm]{ISR53_BPFF1_FIT.eps}
%\includegraphics*[width=6.5cm,height=5.0cm]{LHC7_BPFF1_FIT.eps}
\includegraphics*[width=6.5cm,height=5.0cm]{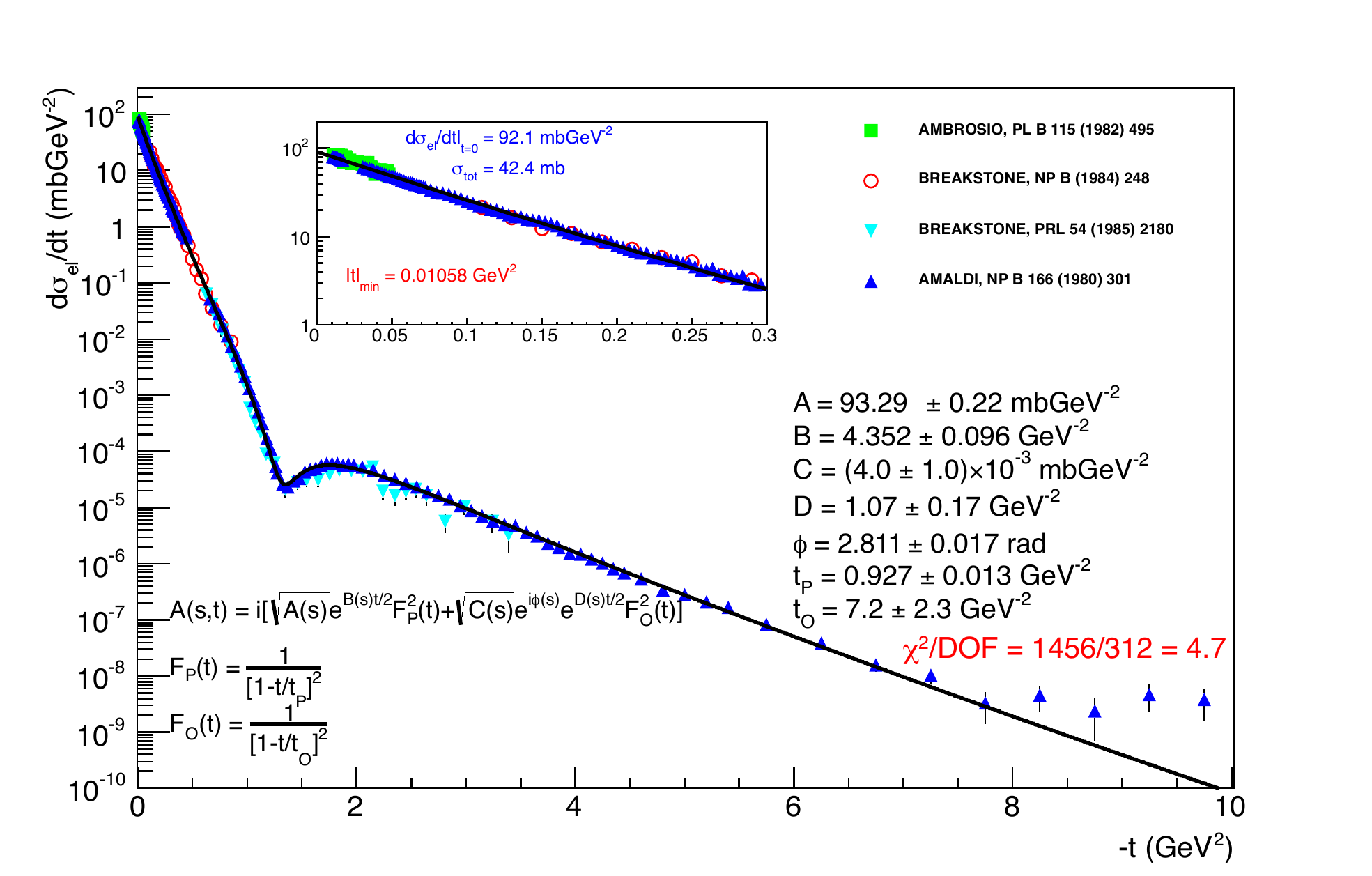}
\includegraphics*[width=6.5cm,height=5.0cm]{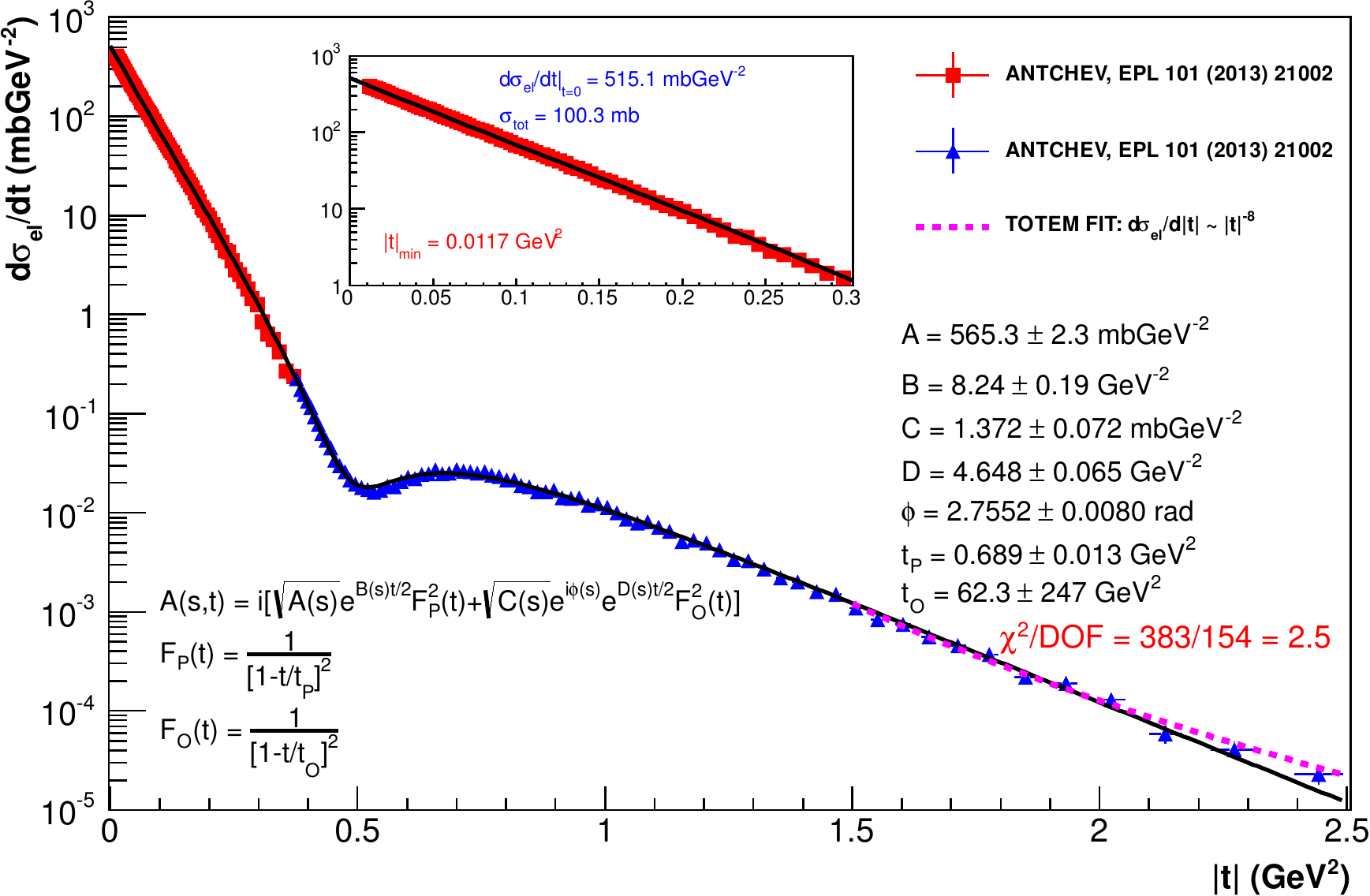}
%}
%LHC7_FIT_oBP_FF_TWODIPOLES}
\caption{Fits to  ISR53 and LHC7 data sets with the two other  modified BP models described { by Eqs.~(\ref{eq:bp6}) and  and (\ref{eq:bp8})}, with ISR fits on the left hand side and LHC fits on the right hand side. Top row: BP amplitude multiplied by an overall { form factor like term}.
% Middle row: only  first term of BP is multiplied by factor.
Bottom row: the two terms in BP amplitude are multiplied by form  factors with different scales. }
\label{g:lhcbp4}
\end{center}
\end{figure}

We show the results of the fit in Fig. ~\ref{g:lhcbp4}.

%Fig05

%FIT_LHC7_mBP2_MAY31-D
An inspection of these fits indicates that an overall multiplicative factor, corresponding to the first top plots, is least favored of the above two possibilities (and less favored than the one chosen in the text, $mBP2$). From the point of view of the $\chi^2$, the fits do not favor  the second possibility relative to the choice $mBP2$, discussed in the text: multiplying both terms by form  factors with different scales or only the first term as in  $mBP2$,  gives an equally good fit, both at ISR and at LHC. However, we notice a problem with  the fits of the bottom figures, when the two terms are each multiplied by a different factor, namely these fits are quite insensitive to the second scale. Phenomenologically therefore, this possibility is not particularly useful, albeit it could be further studied.

\section{Impact parameter structure in the modified models\label{app:impact}}
% [Daniel, May 6th, 2013]\label{app:impact}}
Besides the sum rules, the impact parameter structure of models $mBP1$ and $mBP2$ provides us useful information about unitarity saturation. From our fits with both models, we extract the elastic profile, through the Hankel transform of the amplitude (\ref{eq:bp1}): 
\begin{eqnarray}
\tilde{\mathcal{A}}(s,b) = -i\int_{0}^{\infty} qdqJ_{0}(qb)\mathcal{A}(s,t).\label{eq:hank}
\end{eqnarray}
On the one hand, the dominant contribution comes from the real part, which assume distinct forms for models $mBP1$ and $mBP2$:
\begin{eqnarray}
\tilde{\mathcal{A}}^{mBP1}_{R}(s,b) &=& \sqrt{A}e^{2m_{\pi}\gamma}\mathcal{J}(s,b)+\frac{\sqrt{C}}{D}e^{-b^{2}/2D} \cos\phi; \label{eq:bp12}\\
\mathcal{A}^{mBP2}_{R}(s,b) &=& \sqrt{A}t_{0}^{4}\mathcal{K}(s,b)+\frac{\sqrt{C}}{D}e^{-\frac{b^{2}}{2D}}\cos\phi;\label{eq:bp20}
\end{eqnarray}
where the integrals $\mathcal{J}(s,b)$ and $\mathcal{K}(s,b)$ are given as 
\begin{eqnarray}
\mathcal{J}(s,b) &=& \int_{0}^{\infty} qdqJ_{0}(qb)e^{-Bq^{2}/2-\gamma\sqrt{4m_{\pi}^{2}+q^{2}}};\label{eq:int_J}\\
\mathcal{K}(s,b) &=& \int_{0}^{\infty} qdqJ_{0}(qb)\frac{e^{-Bq^{2}/2}}{(t_{0}+q^{2})^{4}}.\label{eq:int_K}
\end{eqnarray}
On the other, the imaginary part turns out to be the same:
\begin{eqnarray}
\mathcal{A}^{mBP1,mBP2}_{I}(s,b) &=& \frac{\sqrt{C}}{D} e^{-\frac{b^{2}}{2D}}\sin\phi \label{eq:bp21}.
\end{eqnarray}

\begin{figure}[H]
\begin{center}
\includegraphics*[width=8.0cm,height=6.5cm]{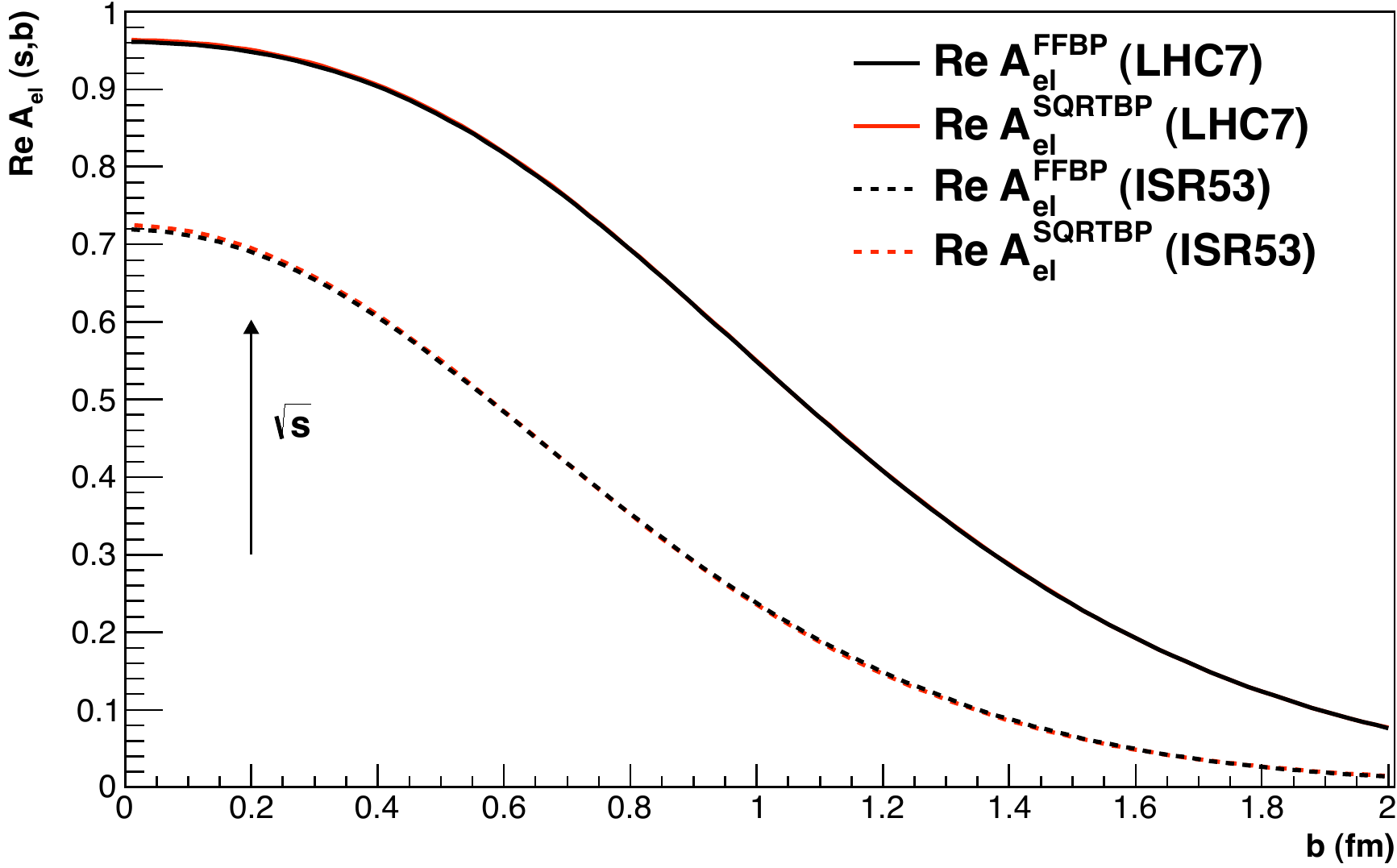}
\includegraphics*[width=8.0cm,height=6.5cm]{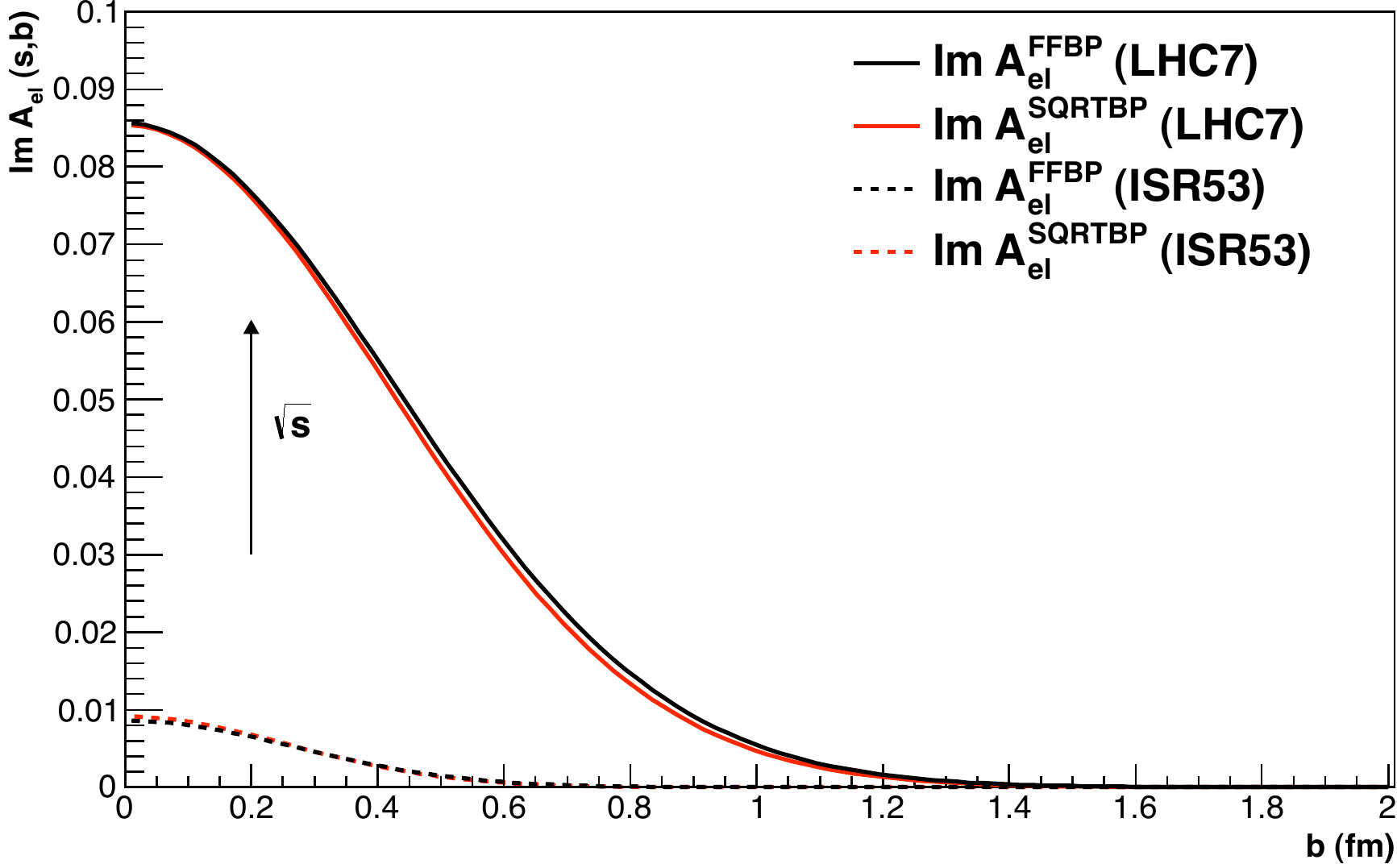}
\caption{Energy evolution of profile functions (real and imaginary) from ISR to LHC7.}
\label{g:lhcbp_profs}
\end{center}
\end{figure}

Unfortunately, due to the introduction of corrections into the first term  of original BP parametrization, the integrals (\ref{eq:int_J}, \ref{eq:int_K}) can no longer be solved analitically. Therefore, we perform numerical evaluations of such integrals. In Fig. \ref{g:lhcbp_profs} we present these calculations and the energy evolution of the elastic $b-$distributions, following from Eqs. (\ref{eq:bp12}-\ref{eq:bp21}), from ISR energies to LHC7. 
 \end{appendices}
 \bibliography{elastic_ARXIV-may31} 
\end{document}